%% file: paper.tex
\documentclass[reprint,
aps,superscriptaddress,twocolumn,showpacs,longbibliography,nofootinbib]{revtex4-2}
\usepackage{graphicx,color}
\usepackage{amsmath,amsfonts,enumerate,amsthm,amssymb,bbm}
\usepackage[colorlinks=true,citecolor=blue,linkcolor=magenta]{hyperref}
\usepackage{multirow}
\usepackage{bbold}
\usepackage{braket}
\usepackage{soul}
\usepackage[caption = false]{subfig}
\usepackage[normalem]{ulem}
\usepackage{scrextend}
\usepackage{xcolor}
\usepackage{dsfont}
\usepackage{nicefrac}
\usepackage[linesnumbered,ruled,vlined]{algorithm2e}
\usepackage{apptools}
\usepackage{physics}
\usepackage[utf8]{inputenc}

\mathchardef\ordinarycolon\mathcode`\:
\mathcode`\:=\string"8000
\begingroup \catcode`\:=\active
  \gdef:{\mathrel{\mathop\ordinarycolon}}
\endgroup

\theoremstyle{plain}
\newtheorem{thm}{Theorem}

\newtheorem{lem}{Lemma}

\theoremstyle{definition}
\newtheorem{defn}{Definition}
\theoremstyle{remark}

\AtAppendix{\counterwithin{thm}{section}}
\AtAppendix{\counterwithin{corol}{section}}
\AtAppendix{\counterwithin{lem}{section}}
\AtAppendix{\counterwithin{propos}{section}}
\AtAppendix{\counterwithin{defn}{section}}
\AtAppendix{\counterwithin{rmk}{section}}
\AtAppendix{\counterwithin{ex}{section}}

\newcommand{\D}{\ensuremath{\mathcal{D}}}
\newcommand{\E}{\ensuremath{\mathbb{E}}}
\renewcommand{\O}{\ensuremath{\mathcal{O}}}
\renewcommand{\S}{\ensuremath{\mathcal{S}}}
\newcommand{\Psiin}{\ensuremath{\ket{\psi_\text{in}}}}

\newcommand{\0}{\ensuremath{\ket{0}}}
\newcommand{\1}{\ensuremath{\ket{1}}}
\newcommand{\+}{\ensuremath{\ket{+}}}
\renewcommand{\-}{\ensuremath{\ket{-}}}
\renewcommand{\i}{\ensuremath{\ket{i}}}
\newcommand{\mi}{\ensuremath{\ket{-i}}}
\newcommand{\kb}[1]{\ensuremath{\ket{#1}\!\bra{#1}}}
\newcommand{\TV}{\text{TV}}
\newcommand{\SWAP}{\text{SWAP}}

\def\<{\langle}
\def\E{ {\mathbb{E}} }

\def\O{ {\cal O} }
\def\D{ {\cal D} }

\def\N{ {\cal N} }

\def\W{ {\cal W} }
\def\I{ \mathbb{1} }

\def\I{ \mathbbm{1} }

\def\>{\rangle}
\def\<{\langle}

\DeclareMathOperator{\poly}{poly}

\renewcommand{\0}{\ensuremath{\ket{0}}}
\renewcommand{\1}{\ensuremath{\ket{1}}}
\renewcommand{\+}{\ensuremath{\ket{+}}}
\renewcommand{\-}{\ensuremath{\ket{-}}}
\renewcommand{\i}{\ensuremath{\ket{i}}}
\renewcommand{\mi}{\ensuremath{\ket{-i}}}

\newcommand{\thv}{\vec{\theta}}

\renewcommand{\norm}[1]{\left\lVert#1\right\rVert}

\renewcommand{\ket}[1]{|#1\rangle}               
\renewcommand{\bra}[1]{\langle #1|}              

\renewcommand{\vec}[1]{\boldsymbol{#1}}  

\newcommand{\stkout}[1]{\ifmmode\text{\sout{\ensuremath{#1}}}\else\sout{#1}\fi}
\newif\ifverbose
\verbosetrue

\usepackage{notes2bib}
\bibnotesetup{note-name=}
\makeatletter 

\renewcommand\onecolumngrid{
\do@columngrid{one}{\@ne}%
\def\set@footnotewidth{\onecolumngrid}
\def\footnoterule{\kern-6pt\hrule width 1.5in\kern6pt}%
}

\begin{document}

\title{The power and limitations of learning quantum dynamics incoherently}

\author{Sofiene Jerbi}
\affiliation{Theoretical Division, Los Alamos National Laboratory, Los Alamos, NM, USA.}
\affiliation{Institute for Theoretical Physics, University of Innsbruck, Austria}

\author{Joe Gibbs}
\affiliation{Department of Physics, University of Surrey, Guildford, GU2 7XH, UK}
\affiliation{AWE, Aldermaston, Reading, RG7 4PR, UK}

\author{Manuel S. Rudolph}
\affiliation{Institute of Physics, Ecole Polytechnique F\'{e}d\'{e}rale de Lausanne (EPFL), CH-1015 Lausanne, Switzerland}

\author{Matthias C.~Caro}
\affiliation{Institute for Quantum Information and Matter, Caltech, Pasadena, CA, USA.}
\affiliation{Dahlem Center for Complex Quantum Systems, Freie Universit\"{a}t Berlin, Berlin, Germany.}

\author{Patrick~J.~Coles} 
\affiliation{Theoretical Division, Los Alamos National Laboratory, Los Alamos, NM, USA.}
\affiliation{Normal Computing Corporation, New York, New York, USA.}

\author{Hsin-Yuan Huang}
\affiliation{Institute for Quantum Information and Matter, Caltech, Pasadena, CA, USA.}
\affiliation{Department of Computing and Mathematical Sciences, Caltech, Pasadena, CA, USA}

\author{Zo\"{e} Holmes}
\affiliation{Theoretical Division, Los Alamos National Laboratory, Los Alamos, NM, USA.}
\affiliation{Institute of Physics, Ecole Polytechnique F\'{e}d\'{e}rale de Lausanne (EPFL), CH-1015 Lausanne, Switzerland}

\date{\today}

\begin{abstract}
Quantum process learning is emerging as an important tool to study quantum systems. While studied extensively in coherent frameworks, where the target and model system can share quantum information, less attention has been paid to whether the dynamics of quantum systems can be learned without the system and target directly interacting. Such incoherent frameworks are practically appealing since they open up methods of transpiling quantum processes between the different physical platforms without the need for technically challenging hybrid entanglement schemes. Here we provide bounds on the sample complexity of learning unitary processes incoherently by analyzing the number of measurements that are required to emulate well-established coherent learning strategies. We prove that if arbitrary measurements are allowed, then any efficiently representable unitary can be efficiently learned within the incoherent framework; however, when restricted to shallow-depth measurements only low-entangling unitaries can be learned. We demonstrate our incoherent learning algorithm for low entangling unitaries by successfully learning a 16-qubit unitary on \texttt{ibmq\_kolkata}, and further demonstrate the scalabilty of our proposed algorithm through extensive numerical experiments.  
\end{abstract}

\maketitle

\section{Introduction}

Classical computing power and classical machine learning are currently routinely used to process data from quantum experiments. However, as these are \textit{quantum} experiments, it would perhaps be more natural to use a \textit{quantum} computer to process their data. 
Within this line of thought, a particularly promising application of quantum hardware is \textit{quantum process learning}~\cite{bisio2010optimal, poland2020no, sharma2020reformulation, khatri2019quantum, Jones2022robustquantum,heya2018variational, cirstoiu2020variational,gibbs2021long, gibbs2022dynamical, huang2021information, huang2021quantum, caro2021generalization, caro2022outofdistribution, huang2022learning, caro2022learning}. At its simplest, the process to be learned will be the unitary dynamics of an experimental system that one wishes to study, and the machine learning model to be optimized will be a quantum circuit representation or a classical description of the target process. In this manner, quantum process learning is a means of digitizing an analogue quantum process and uploading it to a quantum or classical computer for further study. 

\begin{figure}[t]
\centering
\includegraphics[width =\columnwidth]{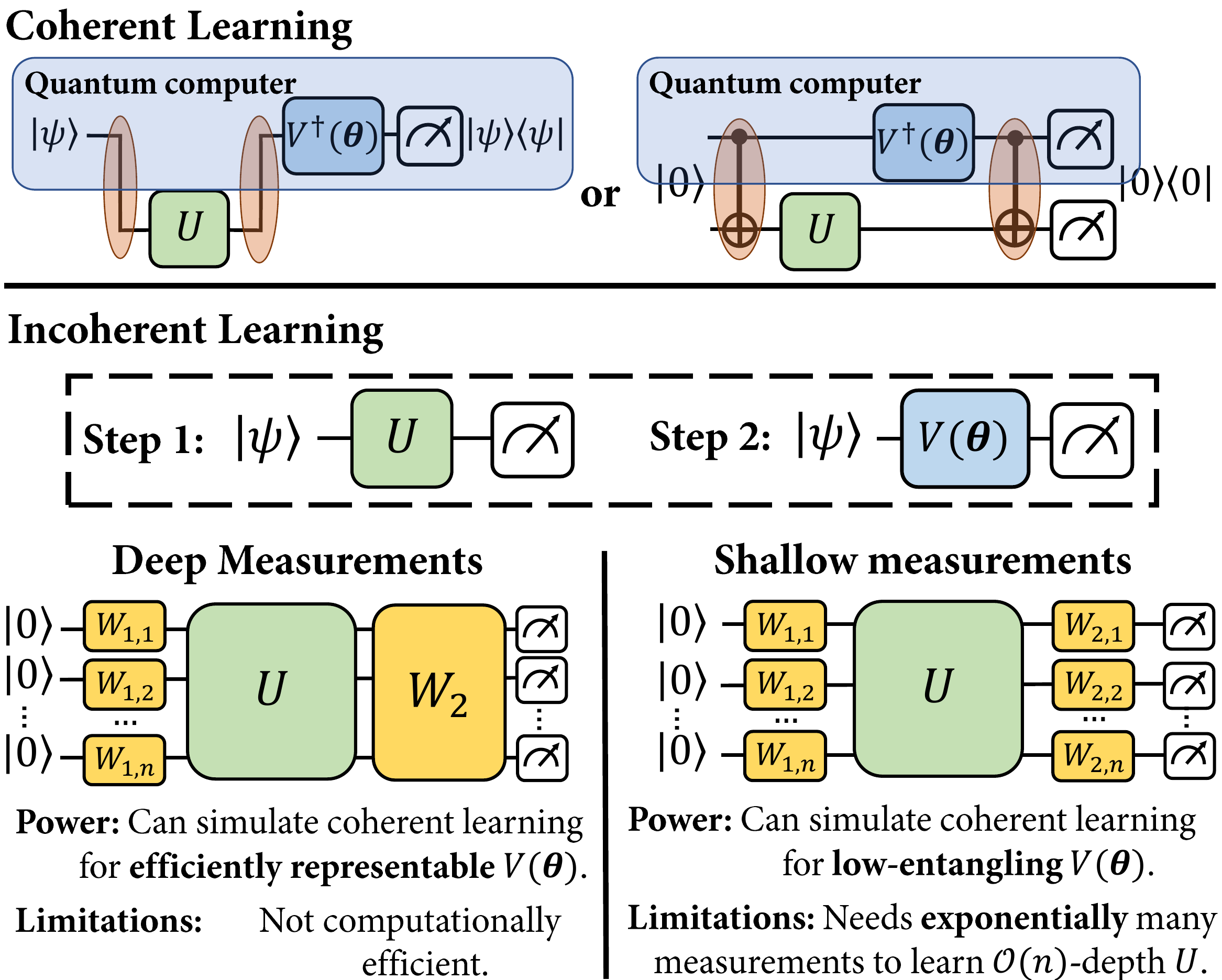}
\vspace{-6mm}
\caption{\textbf{Coherent versus incoherent unitary learning.} We consider the task of learning a target unitary $U$ using a variational ansatz $V(\thv)$. As opposed to a coherent setting where quantum information can be shared or entangled between the quantum computer running $V(\thv)$ and the system implementing $U$, we consider an incoherent setting. This setting is divided into two phases: 1.)~a measurement phase where information is collected about $U$, and 2.)~a training phase where this information is used to train $V(\thv)$.}
\label{fig:schematic}
\vspace{-1.5em}
\end{figure}

Quantum process learning has thus far been predominantly investigated in the \textit{coherent} setting. That is, it is assumed that the target and model system can coherently interact and quantum information can be shared between them. In this setting, the training data typically takes the form of `input-output' pairs of quantum states. 
While the coherent setting is theoretically powerful (in the sense that one can learn efficiently implementable unitaries using only a polynomial number of `easy-to-prepare' product states~\cite{caro2022outofdistribution}), engineering the coherent interaction of the target and model system can be experimentally challenging. In particular, if the target and model system are made up of physically different constituents, then a reliable quantum transducer is required in this framework (see Fig.~\ref{fig:schematic}). However, hybrid-entanglement engineering is still in its infancy~\cite{mao2021perspective, Waks2009Protocol, Eichler2012Observation, Stute2012Tunable, Craddock2019Quantum, scarlino2019coherent} and it may be many years before such transducers are widely available. More fundamentally, given the need for the target and model system to coherently interact, the model system cannot be purely classical.

The \textit{incoherent} setting, where the training is performed using classical data (i.e., numerical values collected from measurement outcomes, see Fig.~\ref{fig:schematic}), is considerably more practical. Crucially, as there is no need for the target system and model system to physically interact in this case, the target system and model system can be radically different. In some cases, namely those where the target unitary is classically simulable, the trained unitary could even be a fully classical model such as those enabled by tensor network methods~\cite{Vidal2003Efficient, orus2014practical}. 
Thus, the incoherent framework provides a way of transpiling quantum processes between quantum and classical platforms. The appeal of this flexibility is reflected in the growing body of literature studying so-called `distributed learning' benchmarking protocols~\cite{anshu2022distributed}, as well as various proposed algorithms for learning Hamiltonians that implicitly assume an incoherent setting~\cite{wang2017experimental, Wiebe2014Quantum, wiebe2014hamiltonian, Gentile2021Learning, wilde2022scalably, gebhart2022learning, stilck-franca2022efficient, haah2021optimal, anshu2021sample-efficient}. 

Recently, the power of the incoherent setting has begun to be investigated from a learning theory perspective. For example, Ref.~\cite{huang2022learning} develops an algorithm for predicting local properties of the output states of an unknown process using only local measurements, and Ref.~\cite{fanizza2022learning} investigates the power of the incoherent setting in the context of hypothesis testing. Here we take a different approach and investigate the extent to which the full coherent setting can be emulated within the incoherent setting in the context of learning unitary processes. 

We start by considering the power of the incoherent paradigm when arbitrary measurements can be performed on the system. In this case, using results from random measurement theory~\cite{huang2020predicting, elben2022randomized} and covering net arguments~\cite{caro2021generalization}, we argue that it is possible to incoherently learn arbitrary efficiently representable unitaries using Clifford shadows and an efficient number of samples. This result highlights that quantum transducers are not strictly always needed for quantum process learning. 

However, there are two crucial caveats to this first result. Firstly, Clifford shadows require the ability to implement circuits of linear depth and so are impractical on many near-term platforms. More fundamentally, there are good reasons to believe (due to the globality of the cost and difficulties surrounding computing it efficiently on quantum or classical hardware) that the algorithm we use for our sampling complexity proof is not computationally efficient in general. As such, we see this algorithm as a predominantly a tool to understand the sampling complexity of the incoherent setting rather than a practical proposal.  

Thus, the core of this paper investigates the more realistic situation where the experimentalist is limited to performing \textit{shallow} measurements, that is, those that can be implemented using shallow depth circuits. In this case, we prove that low-entangling unitaries can be efficiently learned using only Pauli product measurements. However, arbitrary unitaries cannot be learned in this setting even if they are efficiently implementable (that is, even if they can be approximated using a parameterized circuit with a polynomial number of trainable parameters). Our negative result here provides a fundamental limit to quantum process learning in the incoherent paradigm. In particular, contrasting our results with those portrayed in Ref.~\cite{huang2020predicting}, this result highlights that learning to predict the output states of a unitary is a substantially harder learning task than learning to simply predict local measurement outcomes.

In parallel, our positive result for low-entangling unitaries opens up a family of new tools for studying the short-time dynamics of locally interacting systems. We demonstrate the scalability, and plausible computational efficiency, of these tools via numerical simulations for different entangling rates and different system sizes. We further establish the suitability of the algorithm for near-term hardware by successfully learning a 16-qubit unitary on real quantum hardware (\texttt{ibmq\_kolkata}).

\section{Background}

In this work we consider the task of learning to optimize the parameters $\vec{\theta}$ of $V(\vec{\theta})$, an $n$-qubit parameterized unitary, such that, for the optimized parameters $\vec{\theta}_{\rm opt}$, $V(\vec{\theta}_{\rm opt})$ well approximates an unknown $n$-qubit unitary $U\in\mathcal{U}(\mathbb{C}^{2^n})$. The Hilbert-Schmidt inner product between $U$ and $V(\vec{\theta})$ provides a natural measure of the success of the training. Thus our goal is to minimize the Hilbert-Schmidt test cost~\cite{khatri2019quantum},
\begin{equation}
    C_{\text{HST}}(\vec{\theta}) := 1 - \frac{1}{d^2}\abs{\Tr[U^\dagger V(\vec{\theta})]}^2 \, ,
\end{equation}
which can be measured on quantum hardware using Choi states~\cite{khatri2019quantum}. 

The Hilbert-Schmidt cost is operationally meaningful in virtue of its equivalence to the average fidelity between a state $\ket{\psi}$ evolved under $U$ and a state $\ket{\psi}$ evolved under $V(\vec{\theta})$. That is,
\begin{equation}
    C_{\text{HST}}(\vec{\theta}) = \frac{d+1}{d} \E_{\ket{\psi}\sim\text{Haar}_n} \left[1-\abs{\bra{\psi}U^\dagger V(\vec{\theta})\ket{\psi}}^2\right]  \, ,
\end{equation}
where $\ket{\psi}$ is drawn from the (Haar) uniform distribution of states. 
This suggests an alternative approach to training, where the learner has access to a training data set consisting of input-output pairs of pure $n$-qubit states, 
\begin{equation}\label{eq:training-data-general}
    \mathcal{D}_{Q}(N) = \{(\ket{\psi^{(j)}}, U\ket{\psi^{(j)}}) \}_{j=1}^{N} \, ,
\end{equation}
where the $N$ input states $\ket{\psi^{(j)}}$ are drawn independently from the distribution $Q$. In particular, taking $Q$ to be the Haar distribution, $ V(\vec{\theta})$ can be trained using the loss
\begin{equation}\label{eq:GlobalTrainingCost}
    C_{\D_{Q}(N)}(\vec{\theta}) := 1 - \frac{1}{N}\sum_{i=1}^{N} \abs{\bra{\psi^{(i)}}U^\dagger V(\vec{\theta})\ket{\psi^{(i)}}}^2 \, .
\end{equation}
Prior results on \textit{in-distribution} generalization give bounds on the number of training pairs $N$ needed to ensure that this training loss is faithful to the Hilbert-Schmidt cost. In particular, for a variational circuit $V(\vec{\theta})$, with $T$ trainable gates, $N \sim T\log(T)$ training pairs suffice~\cite{caro2021generalization}. 

In fact, more recent results on \textit{out-of-distribution} generalization~\cite{caro2022outofdistribution} imply that simple products of single-qubit stabilizer states suffice to generalize well on globally Haar random states. That is, the training ensemble $Q$ can be taken to be the uniform distribution over the set of stabilizer states $\text{Stab}_1^{\otimes n} := \{\0, \1, \+, \-, \i, \mi\}^{\otimes n}$.
This is practically useful because, in contrast to Haar random states which require deep circuits, a complete gate set, and cannot be classically simulated, random stabilizer product-states are classically simulatable and can be prepared using a single layer of single qubit rotations. 

In practise, to avoid cost-function-dependent barren plateaus~\cite{cerezo2021cost}, it is necessary to use a local variant of the training loss. A natural choice using product state training data is 
\begin{equation}\label{eq:localtrainingloss}
\begin{aligned} \small
        &C_{\rm N}^{\ell}(\vec{\theta}) 
         = 1 -\frac{1}{N}\sum\limits_{j=1}^N \Tr \left[  \ket{\psi^{(j)}}\bra{\psi^{(j)}}  U^\dagger V(\vec{\theta}) H_{\ell}^{(j)} V^\dagger(\vec{\theta}) U\right].
\end{aligned}
\end{equation}
\normalsize
Here $\ket{\psi^{(j)}} = \bigotimes_{i=1}^n \ket{\psi_i^{(j)} } \sim \text{Stab}_1^{\otimes n}$ for all $j$ and $ H_{\ell}^{(j)} = \frac{1}{n} \sum_{i=1}^n \ket{\psi_i^{(j)}} \bra{\psi_i^{(j)} }\otimes \mathds{1}_{\bar{i}}$ is a 1-local measurement (i.e. each term acts non-trivially on at most one qubit). This local cost is faithful to the global product state training cost in the sense that it vanishes under the same conditions~\cite{khatri2019quantum}, but enjoys trainability guarantees for shallow depth circuits~\cite{cerezo2021cost}.
More concretely, in the coherent setting, the following bound holds    
\begin{equation*}
    \begin{aligned}    
        C_{\text{HST}}(\vec{\theta})
        \leq \, &\frac{2 n (d+1)}{d} C^{\ell}_{  N}(\vec{\theta}) + \mathcal{O} \left( n\sqrt{\frac{T \log (T)}{N}} \right)\, , 
    \end{aligned}
    \end{equation*}
with high probability over the choice of product state training data of size $N$~\cite{caro2022outofdistribution}. Thus the local-product-state cost in Eq.~(\ref{eq:localtrainingloss}) can be used to indirectly train the Hilbert-Schmidt Test cost and thereby learn the target unitary $U$.

It is natural to compute the product-state training cost, and its local variants, in the coherent framework. This can be done via a Loschmidt echo type circuit as sketched in Fig.~\ref{fig:schematic} (the top left circuit).
While such circuits are relatively straightforward to run on a single quantum device, they are much harder to implement if the system implementing the target unitary and the platform implementing the trained unitary are distinct. In such cases, experimentally challenging quantum transducers are required to pass the quantum information between the target and learning systems. This motivates considering the incoherent learning paradigm where the target and model systems do not need to interact. The key question tackled in this paper is: 
\emph{Is it possible to emulate the \textit{coherent} learning setting using only \textit{incoherent} learning  protocols?} More precisely, is it possible to rigorously estimate the training losses used coherently, without the target and test system directly interacting?

\section{Results}

\subsection{Deep measurements}

Let us start by considering what can be computed in the incoherent learning paradigm if we suppose that arbitrary measurements can be performed on the system. In this case, we can use random Clifford measurements to generate the Clifford shadow of each of the $N$ output states in our training ensemble~\cite{huang2020predicting}. This allows us to compute the fidelities of our target output states with an exponential number of guess output states (i.e., parameter settings for the parametrized circuit) after collecting a polynomial-sized shadow. Combining this observation with a covering set argument of all efficiently representable unitaries, we show it is possible to train a variational unitary $V(\thv)$ to achieve $C_{\text{HST}}(\vec{\theta}) \leq \epsilon$ for any target accuracy $\epsilon$ with a polynomial sampling complexity. As proven and stated more formally in Appendix~\ref{app:DeepMeasurements}, this is captured by the following theorem \bibnote{We note that a similar argument was made in passing in Ref.~\cite{caro2022outofdistribution}, but the argument was not formalized and its implications for incoherent learning were not appreciated.}.

\begin{thm}[Power of incoherent learning with deep measurements (informal)]\label{thm:powerofdeep}
For any efficiently representable $n$-qubit unitary $U$, that is a unitary that can be implemented using a quantum circuit of size $\poly(n)$, at most $\poly(n)$ calls to $U$ are required to incoherently train a parameterized unitary $V(\vec{\theta})$ to achieve $1/\poly(n)$ Hilbert-Schmidt cost between $V(\vec{\theta})$ and $U$. 
\end{thm}

In showing this theorem, we also prove that, if random Clifford measurements on the target system are allowed, then it is possible to fully emulate the coherent setting. That is, it is possible to efficiently (in terms of sampling complexity) train $V(\vec{\theta})$ to learn $U$ in the incoherent setting.
It follows that if the target system allows for a Clifford gate set and long coherence times (all $n$-qubit Clifford operations can be implemented in depth $\O(n)$ \cite{maslov2018shorter}) then quantum transducers are not strictly needed for quantum process learning.

However, there are two important caveats here. Firstly, Clifford shadows are only well-suited to compute global fidelities and therefore a global cost (of the form of Eq.~(\ref{eq:GlobalTrainingCost})). However, global costs exhibit barren plateaus. Thus, using generic ans\"{a}tze, the time complexity of this approach will typically be exponential~\cite{cerezo2021cost}. One way around this would be to build prior knowledge about the target unitary into the guess ansatz. It has been shown, for example, that highly symmetrized ans\"{a}tze can avoid globality induced barren plateaus~\cite{schatzki2022theoretical}. Nonetheless, such approaches are unlikely to work in all cases. Secondly, while Clifford shadows are guaranteed information-theoretically to allow the computation of fidelities with other quantum states, this computation is by no means guaranteed to be efficient. When computing fidelities using Clifford shadows, we are indeed considering observables of the form $V(\thv)\ket{\psi^{(i)}}\bra{\psi^{(i)}}V^\dagger(\thv)$, which cannot be efficiently represented/computed classically for deep circuits $V(\thv)$. Even when we consider access to a quantum computer to compute the overlap between these operators, we face the problem that classical shadows are completely unphysical operators that cannot be efficiently encoded into quantum states and whose expectation values are costly to estimate (see Appendix~\ref{app:ComputComplexIncoherent} for an in-depth discussion).

These two constraints limit the breadth of applicability of Theorem \ref{thm:powerofdeep}. Hence, we will now turn our attention to the focus of this work: the power and limitation of the incoherent framework if we limit ourselves to shallow measurements. 

\subsection{Shallow measurements}

\begin{figure}[t]
\centering
\includegraphics[width =\columnwidth]{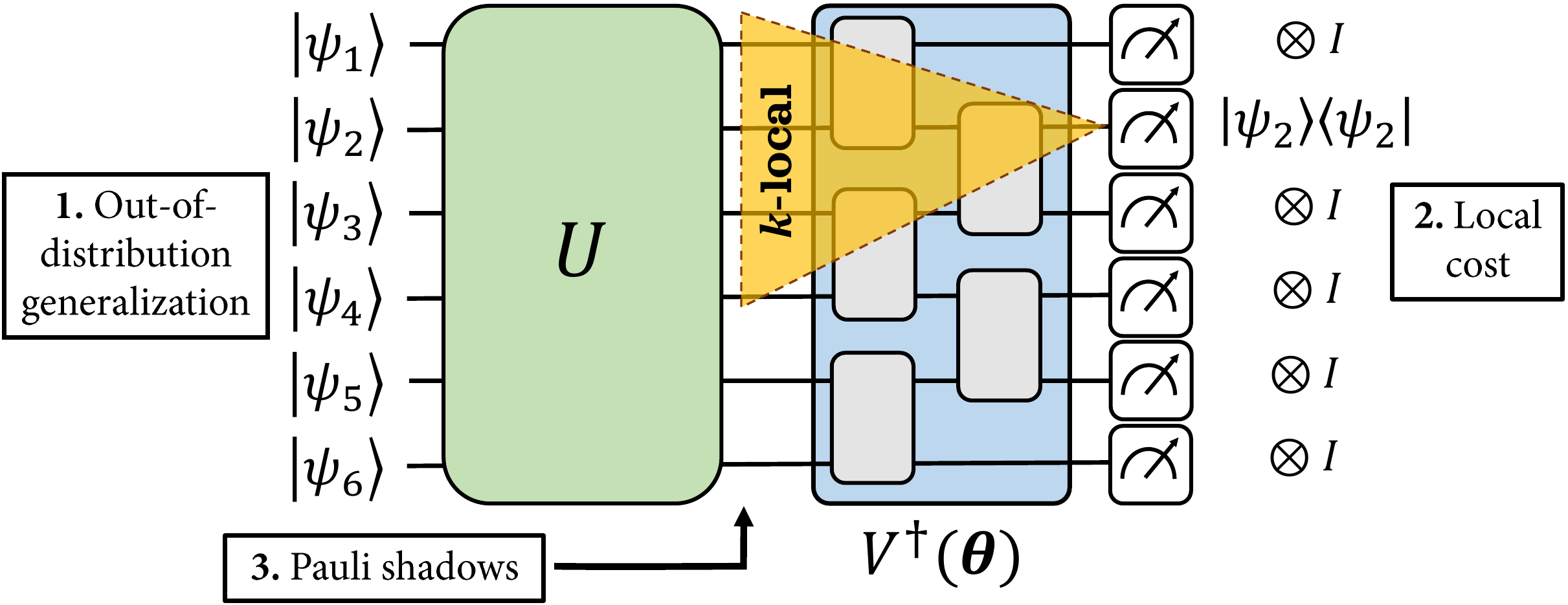}
\vspace{-6mm}
\caption{\textbf{Learning low-entangling unitaries using Pauli shadows.} Our learning protocol for low-entangling unitaries makes use of three results. 1.~Out-of-distribution generalization allows to consider locally-scrambled training states with generalization guarantees to global Haar random states. 2.~Cost concentration results allow us to consider a local version of the training loss. 3.~For shallow-depth ans\"{a}tze $V(\thv)$ (e.g., depth $\O(\log(n))$), the observables appearing in the training loss remain local ($k\in\O(\log(n))$), and therefore Pauli shadow allow to efficiently simulate this training loss incoherently.}
\label{fig:lightcone}
\vspace{-3mm}
\end{figure}

We can follow a similar approach as above with Clifford shadows, but this time restrict ourselves to Pauli shadows. In this case, since the sample complexity of Pauli shadows scales exponentially with the locality of the operators measured, we can only efficiently compute the effect of measuring operators that scale at most logarithmically with the size of the system. It follows that we can only emulate the coherent setting and estimate $C_{N}^\ell(\thv)$ for low-entangling unitaries that transform $1$-local operators into $O(\log(n))$-local operators, as sketched in Fig.~\ref{fig:lightcone} and further detailed in Appendix~\ref{app:ShallowMeasurements}. Thus, we show that the following theorem holds. 

\begin{thm}[Power of incoherent learning with shallow measurements (informal)]\label{thm:powerofshallow}
For any \textit{low entangling} $n$-qubit unitary $U$, that is a unitary that transforms $1$-local operators into $O(\log(n))$-local operators, at most $\poly(n)$ calls to $U$ are required to incoherently train a parameterized unitary $V(\vec{\theta})$ to achieve $1/\poly(n)$ Hilbert-Schmidt cost between $V(\vec{\theta})$ and $U$. 
\end{thm}

As opposed to the deep-measurement protocol, the shallow-measurement protocol is also computationally efficient in simulating the coherent training loss $C_{\rm N}^{\ell}(\vec{\theta})$ (see Eq.~(\ref{eq:localtrainingloss})). Since the observables $V(\vec{\theta}) H_{\ell}^{(j)} V^\dagger(\vec{\theta})$ are $\O(\log(n))$-local, they can be represented and computed efficiently classically, along with their contribution to the training loss. This allows one to train $V(\thv)$ classically (see Appendix~\ref{app:ComputComplexIncoherent} for an in-depth discussion) and thereby learn a classical representation of the target process. If $V(\thv)$ is parameterized such that it can be implemented on a different quantum system, this approach further provides a way of transpiling the quantum process between (potentially rather different) quantum platforms. 
 
The Pauli shadow based approach for learning low entangling unitaries via shallow measurements in the incoherent framework breaks down if the target unitary is too entangling as the sample complexity of our method scales as $4^k$ where $k$ is the locality of the operator $V(\vec{\theta}) H_\ell V(\vec{\theta})^\dagger$. It turns out that this is not a limitation of our algorithm but rather a fundamental limitation of incoherent learning with shallow measurements. In particular, we prove that there exist efficiently implementable unitaries such that an exponential sample complexity is required to learn them in the incoherent framework with only single qubit measurements. This is captured by the following theorem. 

\begin{figure}[t]
\centering
\includegraphics[width =\columnwidth]{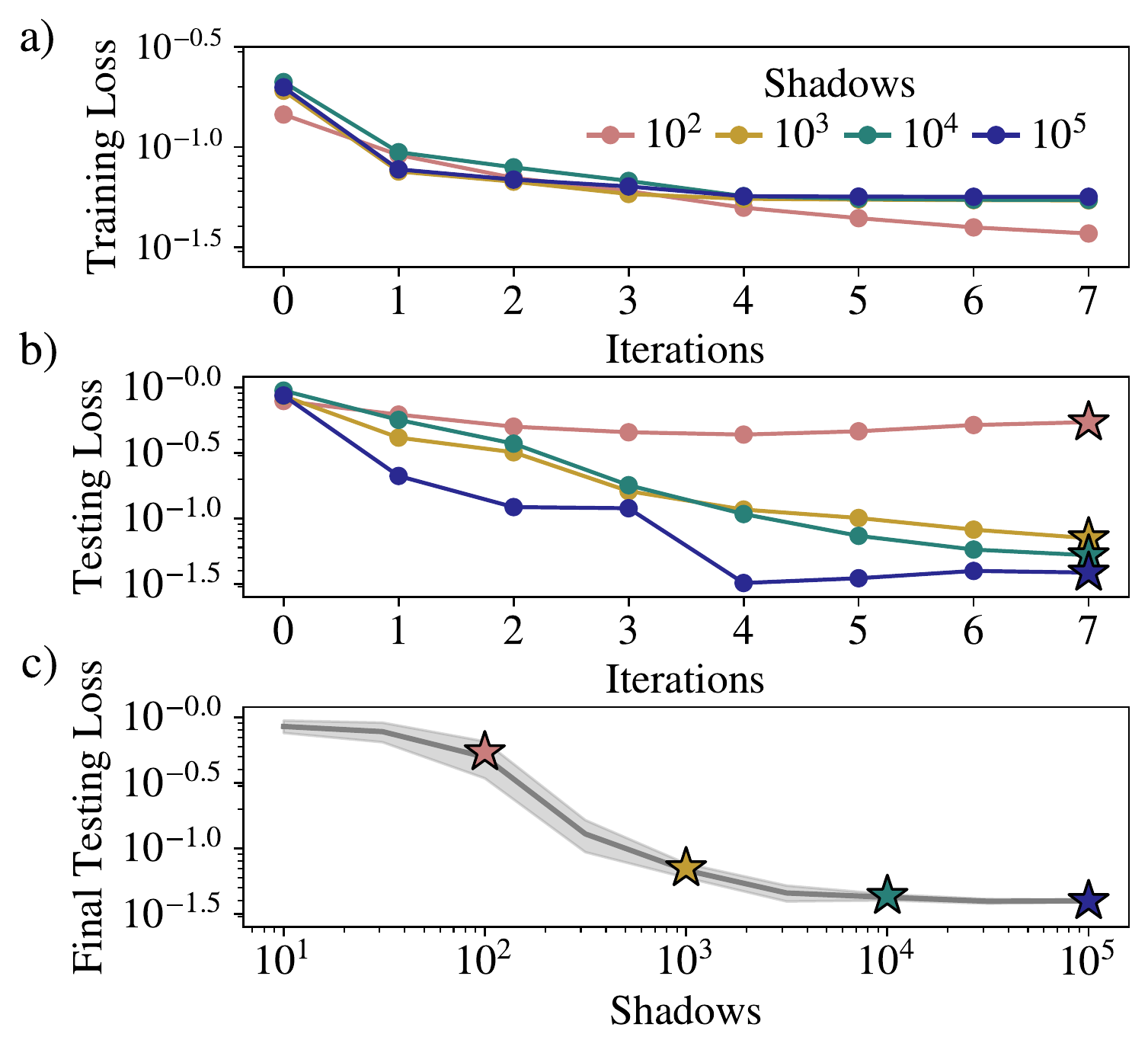}
\vspace{-8mm}
\caption{\textbf{Incoherent Learning of 16-qubit Dynamics on Quantum Hardware.}  A short-time Trotterisation of a 16-qubit Transverse-Field Ising Model implemented on the \texttt{ibmq\_kolkata} device is incoherently compiled using shallow measurements.  The PQC's rotation angles are initialized from the uniform distribution $\theta_i \sim \mathcal{U}(-1, 1)$. a) and b) shows typical training and testing curves. The errors decrease as the optimization progresses, indicating the target unitary is being successfully compiled. In c), the testing error upon convergence is shown to decrease for an increasing number of shadows trained on. Here the mean and standard deviation shown here are produced from 10 repeats of training on independent sets of shadows. The star markers emphasize that the testing losses plotted in c) are produced from the statistics of the converged final testing losses of repeated optimizations.}
\label{fig:Hardware_plot}
\vspace{-3mm}
\end{figure}

\begin{figure*}[t!]
\centering
\includegraphics[width =\textwidth]{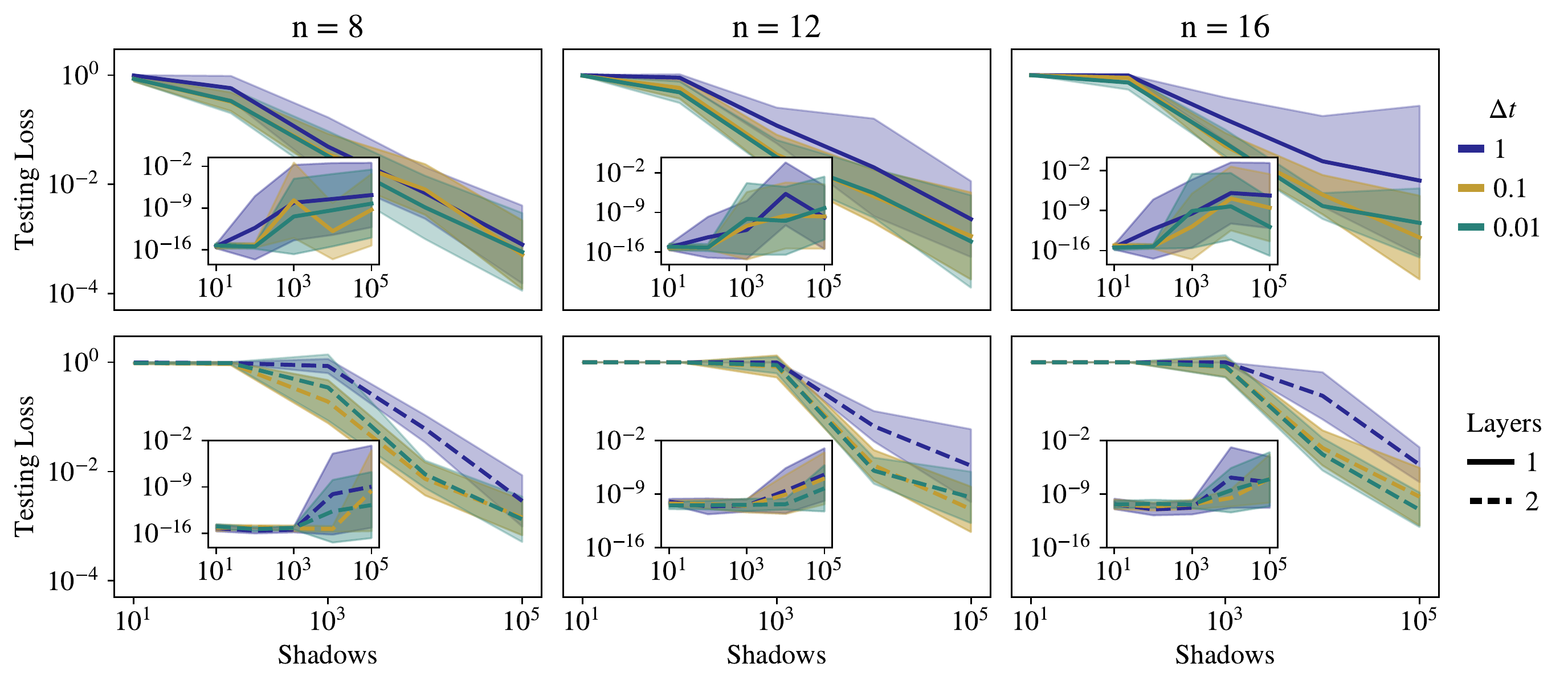}
\vspace{-6mm}
\caption{\textbf{Scalability of Incoherent Learning.} For a range of system sizes ($n$), shadow sizes ($M$), Trotter timesteps ($\Delta t$) and number of ansatz layers, we perform the Incoherent Learning algorithm using shallow measurements. For each configuration, we save the final testing loss (a product state overlap approximation of the Hilbert-Schmidt Test), and 10 repeats generate the mean and standard deviation plotted. The insets show the corresponding training loss upon termination of the optimization.}
\label{fig:Simulation_plot}
\end{figure*}

\begin{thm}[Limit of incoherent learning with shallow measurements (informal)]\label{thm:limitofshallow}
There exist efficiently representable unitaries $U$ such that when restricted to performing only shallow measurements, at least $2^{\Omega(n)}$ calls to $U$ are required to achieve a Hilbert-Schmidt cost $C_\textnormal{HST}(\thv)\leq 1/4$ in the incoherent framework. 
\end{thm}

As discussed in more detail in Appendix \ref{app:lower-bounds-shallow}, we prove this theorem by reduction of the task of distinguishing between orthogonal unitaries that can be used to prepare GHZ-like states. We show that the distinguishing task requires exponential sample complexity for shallow measurements. Moreover, the unitaries we consider are implementable using a linear depth circuit (see Fig.~\ref{fig:GHZ-unitaries} in the Appendix). It follows that there exist efficiently representable unitaries that cannot be efficiently learned using Pauli- (or more general product-) measurements in the incoherent setting. Theorem~\ref{thm:limitofshallow} amounts to a fundamental no-go theorem for the incoherent learning framework and prohibits the emulation of the coherent framework using single qubit Pauli measurements~\cite{huang2021information, chen2021exponential, huang2021quantum, aharonov2022quantum, caro2022learning}. 

\subsection{Implementations}

\paragraph*{Quantum Hardware.}

We demonstrate the ability of our shallow-measurement incoherent-learning algorithm to learn a 16-qubit low-entangling unitary on \texttt{ibmq\_kolkata}. Specifically, the target unitary is a first-order short-time ($\Delta t = 0.1$) Trotterisation for the Transverse-Field Ising Model with Hamiltonian 
\begin{equation}
    H_{\rm Ising} = \sum_{i=0}^{14} Z_i Z_{i+1} + \sum_{i=0}^{15}\alpha_i X_i \, ,
\end{equation} where $\alpha_i$ are randomly drawn from the normal distribution $\mathcal{N}(\mu\,{=}\,0, \sigma\,{=}\,0.5)$. As described in detail in Appendix~\ref{app:ShallowMeasurements}, a Pauli classical shadow of size $280000$ is created for only two training states $\{ U(\Delta t)|\psi_i\rangle \}_{i=0}^1$, where $|\psi_i\rangle$ is a Haar random product state. (Random product states enjoy the same out-of-distribution generalization guarantees as products of random single qubit stabilizer states~\cite{caro2022outofdistribution}).
These shadows are used to compute the local training cost, which is minimised classically by training a simulated parameterized quantum circuit (PQC) with the same gate structure as the target Trotterised unitary. We compute the testing loss $C_{\mathcal{D}_{\text{Haar}_1^{\otimes n}(N)}}(\thv)$ for a set of $N=100$ random product states. This coherent product-state overlap is an operationally meaningful quantity, since it correlates, up to a factor of 2 and additive error $\mathcal{O}\Big(\sqrt{\frac{1}{N}}\Big)$, with the Hilbert-Schmidt cost between the target and learned unitaries, $C_{\text{HST}}(\vec{\theta})$~\cite{caro2022outofdistribution}.

\medskip

Fig.~\ref{fig:Hardware_plot}a) shows the training loss decreasing as the L-BFGS optimizer~\cite{liu1989limited} minimises the cost function, Fig.~\ref{fig:Hardware_plot}b) shows the corresponding testing loss decreasing and Fig.~\ref{fig:Hardware_plot}c) shows the final testing error as a function of shadow size. The differed sized shadows used here are created from randomly chosen subsets of the full 280000 sample shadow originally collected on the quantum computer. For small shadow sizes the final testing loss decreases (signifying improved generalization) as the shadow size increases, but for shadows above $\sim 10^4$ the improvement to the final testing error plateaus. This indicates that a shadow of size $\sim 10^4$ suffices to well reproduce the true underlying output states.

\medskip

\paragraph*{Simulations.}
We now investigate the scalability of the incoherent-learning algorithm with classical simulations of the technique. Here we focus on the $n$-qubit Heisenberg Hamiltonian with open boundary conditions,
\begin{equation}
    H_{\rm Heis}
     = \sum_{i=0}^{n-1} X_i X_{i+1}+Y_i Y_{i+1}+Z_i Z_{i+1} \, .
\end{equation}
We learn a first order (1-layer) and second order (2-layer) Trotterisation of $H_{\rm Heis}$ using an ansatz of the same structure and using two training states $\{ U(\Delta t)|\psi_i\rangle \}_{i=0}^1$ where $|\psi_i\rangle$ are product states. Fig.~\ref{fig:Simulation_plot} compares the final testing loss found for a range of system sizes, shadow sizes and number of layers in the ansatz. Similarly to the hardware results in Fig.~\ref{fig:Hardware_plot}, the testing error is measured by an approximation to the Hilbert-Schmidt Test, computed using product state overlaps. 

We find that for increasing system sizes, there is not a significant decline in the testing performance. Conversely, as the degree of entanglement in the target unitary and ansatz increases, by increasing $\Delta t$ and the number of layers, the testing performance degrades. This can be explained by the growth of the light-cone of the local measurement operators under $V^\dagger(\vec{\theta})$ as sketched in Fig.~\ref{fig:lightcone}. Interestingly, the shadow size required for training depends more strongly on $\Delta t$ than on the number of layers, indicating that it is not the absolute size of the light cone, but rather its effective size (namely the region on which the back-propagated local operator has non-negligible weight) that matters. We discuss this phenomenon further in Appendix~\ref{ap:approxlocal}.

\section{Discussion}

In this paper we have established a fundamental limitation on our ability to learn a target process without coherent interaction between the target and guess unitaries. Namely, Theorem~\ref{thm:limitofshallow} shows that there exist efficiently representable unitaries such that exponentially many shallow measurements (e.g., Pauli measurements) are required to emulate the coherent setting. Crucially, our metric for success here is an average fidelity between the true and learned output states. This is to be contrasted with Ref.~\cite{huang2020predicting} which presents an efficient (both in terms of sampling and computational complexity) incoherent learning algorithm for predicting local measurements outcomes on the outputs of arbitrary processes. Thus our no-go theorem highlights that learning to predict the output states of a process, and thereby fully emulate the coherent setting, is a substantially harder learning task than learning to predict local measurement outcomes.

This makes it especially noteworthy that it is possible, as demonstrated by our proposed algorithm, to fully emulate the coherent setting in the case of low-entangling unitaries. Crucially, our algorithm uses only easy-to-prepare product training states and simple Pauli measurements, making it NISQ friendly~\cite{preskill2018quantum}. We have proven (Theorem~\ref{thm:powerofshallow}) that the sample complexity of this approach scales polynomially with the size of the target unitary. While the computational complexity of this algorithm is unknown, we have good reasons to expect it to be efficient, since it uses a local cost which can be efficiently computed and which enjoys trainability guarantees~\cite{cerezo2021cost}.
This optimism is further supported by our successful implementation of the algorithm to learn a 16-qubit physical process from real quantum hardware.

Finally, it is worth mentioning that both here and in Ref.~\cite{huang2020predicting}, as well as numerous other papers on quantum process learning~\cite{khatri2019quantum, cirstoiu2020variational, caro2022outofdistribution, mizuta2022local, younisqfast2021, patelquest2022}, it is tacitly assumed that we are interested in learning the output of a quantum process over randomly chosen input states (or equivalently we are interested in maximizing the Hilbert-Schmidt inner product). However, in practise one may well be interested in learning to predict the output of a quantum process for a given set of physically interesting input states~\cite{caro2022learning}. Generalizing our results to this setting is an important direction for future work. Other potential extensions include the generalization to quantum channels and/or continuous variable systems.\\

\begin{acknowledgments}
    SJ was supported by the U.S. Department of Energy (DOE) through a quantum computing program sponsored by the Los Alamos National Laboratory (LANL) Information Science and Technology Institute. SJ acknowledges support from the Austrian Science Fund (FWF) through the projects DK-ALM:W1259-N27 and SFB BeyondC F7102. SJ also acknowledges the Austrian Academy of Sciences as a recipient of the DOC Fellowship. 
    MCC was supported by a DAAD PRIME fellowship. PJC acknowledges initial support from the Los Alamos National Laboratory (LANL) ASC Beyond Moore's Law project and subsequent support by the U.S. DOE, Office of Science, Office of Advanced Scientific Computing Research, under the Accelerated Research in Quantum Computing (ARQC) program. HH is supported by a Google PhD Fellowship. ZH acknowledges initial support from the LANL Mark Kac Fellowship and subsequent support from the Sandoz Family Foundation-Monique de Meuron
program for Academic Promotion.
\end{acknowledgments}

\bibliography{quantum}

\clearpage 
\input{Supplementary}

\end{document}

%% file: supplementary.tex
\appendix

\setcounter{page}{1}
\renewcommand\thefigure{\thesection\arabic{figure}}
\setcounter{figure}{0} 

\onecolumngrid

\begin{center}
\large{ Supplementary Material for \\ ``The power and limitations of learning quantum dynamics incoherently''
}
\end{center}

\section{Preliminaries}\label{app:Prelim}

\medskip 

\setcounter{thm}{0}
\setcounter{corol}{0}
\setcounter{lem}{0}
\setcounter{propos}{0}
\setcounter{defn}{0}
\setcounter{rmk}{0}
\setcounter{ex}{0}

We consider a unitary compilation scenario where the training data about the target unitary $U$ is not given coherently as $\{\Psiin, U\Psiin\}_{\Psiin}$, but rather as incoherent (classical) data $\{\Psiin, m\}_{(\Psiin,F)}$, where $m$ is the (random) outcome of measuring $\Psiin$ using the POVM $F=\{w_m 2^n \kb{\phi_m}\}_m$, and where $\Psiin$ is given by an efficient classical description. We investigate the power and limitations of different choices of input state distributions and POVMs.

\section{The power of incoherent learning}

We propose algorithms for learning an $n$-qubit unitary $U$ using incoherent data. In this section, we describe how to generate this incoherent data, how to use it to train a parametrized ansatz $V(\thv)$, and study the sample complexity of our proposed algorithms in order to achieve a certain learning performance.

\subsection{Summary of the algorithms}

In order to learn a target unitary $U$, the parametrized ansatz $V(\thv)$ is trained via optimization of a training loss that is representative of the overlap between $U$ and $V(\thv)$. For reasons explained below, we consider different training losses for the incoherent setting using deep measurements and the setting using shallow measurements. In the deep measurement setting we consider the training loss:
\begin{equation}\label{eq:global-training}
    C_{\D(N)}(\vec{\theta}) = 1 - \frac{1}{N}\sum_{i=1}^{N} \abs{\bra{\psi^{(i)}}U^\dagger V(\vec{\theta})\ket{\psi^{(i)}}}^2 ,
\end{equation}
where, for all $j = 1, \ldots, N$, $\ket{\psi^{(j)}} = \bigotimes_{i=1}^n \ket{\psi_i^{(j)} } \sim \text{Stab}_1^{\otimes n}$ is a tensor product of random single-qubit stabilizer states. As for the shallow measurement setting, we look into the local version of this training loss:
\begin{equation}\label{eq:local-training}
        C_{\rm N}^{\ell}(\vec{\theta}) 
         = 1 -\frac{1}{N}\sum\limits_{j=1}^N \Tr \left[  \ket{\psi^{(j)}}\bra{\psi^{(j)}}  U^\dagger V(\vec{\theta}) H_{\ell}^{(j)} V(\vec{\theta})^\dagger U\right], 
\end{equation}
where $H_{\ell}^{(j)} = \frac{1}{n} \sum_{i=1}^n \ket{\psi_i^{(j)}} \bra{\psi_i^{(j)} }\otimes \mathds{1}_{\bar{i}}$.

The goal of our incoherent learning algorithm is to gather incoherent data that can be used to ``simulate'' a coherent learning procedure. More precisely, we want to incoherently estimate the training losses above to a desirable precision $\varepsilon > 0$, for all parameter assignments $\thv$. This then allows us to train the parameters $\thv$ similarly to a coherent setting.\\

\paragraph*{Deep measurement algorithm.} For each of the $N$ training states $U \ket{\psi^{(j)}}$, we first gather an $M$-shot Clifford shadow. That is, we use $M$ copies of the state $U \ket{\psi^{(j)}}$ and, on each copy indexed by $m = 1,\ldots, M$, we apply a randomly chosen Clifford unitary $W_m^{(j)}$ and measure in the computational basis. On doing so, we obtain a basis state $\ket{b_m^{(j)}}$ with probability $\abs{\bra{b_{m}^{(j)}}W_m^{(j)} U\ket{\psi^{(j)}}}^2$. The $M$-shot Clifford shadow \cite{huang2020predicting} of the state $U \ket{\psi^{(j)}}$ is then given by the collection $\{\rho_m^{{(j)}}\}_{m=1}^M$, for
\begin{equation}\label{eq:shadowClifford}
    \rho_m^{{(j)}} = (2^n + 1)  W_m^\dagger|b_m^{(j)}\rangle \langle b_m^{(j)} | W_m - \I .
\end{equation}
From here, no more interaction with the target system is required. Clifford shadows are particularly suited to estimate global observables with support on a small subspace (i.e., observables $O$ such that $\Tr[O^2]$ is small). To see how this makes them interesting for estimating the training loss $C_{\D(N)}(\vec{\theta})$ defined in Eq.~(\ref{eq:global-training}), we re-write it as:
\begin{equation}
    C_{\D(N)}(\vec{\theta}) = 1 - \frac{1}{N}\sum_{i=1}^{N} \Tr[U\ket{\psi^{(j)}}\bra{\psi^{(j)}}U^\dagger\ O^{(j)}(\thv)] ,
\end{equation}
where we have absorbed the parametrized unitary $V(\thv)$ into the observable
\begin{equation}
    O^{(j)}(\thv) = V(\thv) \ket{\psi^{(j)}}\bra{\psi^{(j)}} V^\dagger(\thv).
\end{equation}
Since $\Tr[O^{(j)}(\thv)^2] = 1$, we can therefore use the Clifford shadows of the states $U \ket{\psi^{(j)}}$ to estimate $C_{\D(N)}(\vec{\theta})$.\\

\paragraph*{Shallow measurement algorithm.} In the shallow measurement setting, we instead gather $M$-shot Pauli shadows of the training states. That is, we measure each copy in a random Pauli basis by applying randomly chosen single-qubit Clifford gates on each qubit $W_m^{(j)} = W_{m,1}^{(j)} \otimes \ldots \otimes W_{m,n}^{(j)}$ and measuring in the computational basis. The $M$-shot Pauli shadow \cite{huang2020predicting} of the state $U \ket{\psi^{(j)}}$ is now given by the collection $\{\rho_m^{{(j)}}\}_{m=1}^M$, for
\begin{equation}\label{eq:shadowPauli}
    \rho_m^{{(j)}} = \bigotimes_{i=1}^n \left( 3W_{m,i}^\dagger|b_m^{(j)}\rangle \langle b_m^{(j)} | W_{m,i} - \I \right) .
\end{equation}
From here, no more interaction with the target system is required. Pauli shadows are particularly suited to estimate local observables (i.e., that act non-trivially on a small number of qubits). To see how this makes them interesting for estimating the training loss $C_{\rm N}^{\ell}(\vec{\theta})$ defined in Eq.~(\ref{eq:local-training}), we re-write it as:
\begin{equation}
        C_{\rm N}^{\ell}(\vec{\theta}) 
         = 1 -\frac{1}{nN}\sum\limits_{j=1}^N\sum_{i=1}^n \Tr \left[U \ket{\psi^{(j)}}\bra{\psi^{(j)}}U^\dagger \ O^{(j)}_i(\thv) \right], 
\end{equation}
where we have absorbed the parametrized unitary $V(\thv)$ into the observable
\begin{equation}\label{eq:paramobservables}
    O^{(j)}_i(\thv) = V(\thv) \left(\ket{\psi_i^{(j)}} \bra{\psi_i^{(j)} }\otimes \mathds{1}_{\bar{i}}\right) V^\dagger(\thv).
\end{equation}
The size of the Pauli shadow $M$ required to provide precise estimates for these expectation values scales exponentially with the locality of the observables $O^{(j)}_i(\thv)$. Therefore, for parametrized unitaries that are low-entangling, in the sense that they transform $1$-local operators into $\O(\log(n))$-local operators, $O^{(j)}_i(\thv)$ is then a $\O(\log(n))$-local observable, and the number of copies $M$ required should remain polynomial in $n$.\\

In the following, we analyze in more detail the sample complexity of both these algorithms.

\subsection{Sample complexity analysis}

We divide the analysis of the sample complexity of our incoherent learning protocols into two stages: (i) we analyze the sample complexity to incoherently estimate the training loss $C(\thv)$, via the algorithms described above, for any sequence of parameter settings $\thv^{(1)}, \ldots, \thv^{(L)}$, and (ii) we show how this ability to estimate the training loss can be leveraged to learn an arbitrary target unitary $U$. 
Interestingly, we find in (i) that, for both the shallow and deep measurement settings, this sample complexity scales logarithmically with $L$, the number of parameter settings to be evaluated. 
This allows us in (ii) to perform an exhaustive search for the parameter setting $\thv_\textnormal{opt}$ that approximately minimizes the Hilbert-Schmidt cost between $U$ and $V(\thv)$. But, more generally, the result (i) can be seen as a guarantee that the sample complexity needed to simulate a coherent learning approach, e.g., based on gradient-based optimization of $\thv$, depends at most logarithmically on the number of optimization steps.

Along with the theoretical guarantees of classical shadow estimation \cite{huang2020predicting}, the most important tool that we make use of in our sample complexity analysis is the notion of covering net of a parametrized quantum circuit, and an upper bound on its size \cite{caro2021generalization}. These tools are presented formally in the next definitions and lemmas.

\begin{defn}[Covering net of a parametrized quantum circuit]
Let $V(\thv)$ be a parametrized $n$-qubit quantum circuit. Let $\epsilon > 0$ and $D(U,V)$ a distance measure between unitaries. We call $\mathcal{N}_\epsilon$ an $\epsilon$-covering net of the variational family $\{V(\thv)\}_{\thv}$ an ensemble of unitaries $\{ V(\thv^{(i)})\}_{i=1}^{|\mathcal{N}_\epsilon|}$, such for any $V(\thv)$ accessible to the parametrized quantum circuit,
 \begin{equation}
        \exists i\in\{1,\ldots, |\mathcal{N}_\epsilon|\} \, \, \,  \text{such that} \, \, \, D(V(\thv^{(i)}),V(\thv)) \leq \epsilon.
\end{equation}
\end{defn}

\noindent For our purposes, we only require a covering net for arbitrary 2-qubit unitaries:

\begin{lem}[Covering net for 2-qubit unitaries (Lemma C.1 in \cite{caro2021generalization})]\label{lem:covernet}
For any $\epsilon \in (0, 1]$, the set of 2-qubit unitaries admits an $\epsilon$-covering net $\mathcal{N}_\epsilon = \{ U_i\}_{i=1}^{|\mathcal{N}_\epsilon|}$ with respect to the spectral distance $||U - V ||_{\infty}$ with a size $|\mathcal{N}_\epsilon|$ that satisfies
 \begin{equation}
       |\mathcal{N}_\epsilon| \leq \left(\frac{6}{\epsilon}\right)^{32}.
\end{equation}
\end{lem}

\noindent The sample complexities of Clifford and Pauli shadows are given by:

\begin{lem}[Sample complexity of Clifford shadows (Theorem S1, Example 1 in \cite{huang2020predicting})]\label{lem:sampClifford}
Let $O_1, \ldots, O_M$ be a collection of $M$ Hermitian operators and $\epsilon, \delta' \in [0,1]$ be accuracy parameters. Clifford shadows using $N'$ copies of a quantum state $\rho$ allow to compute estimates $\hat{o}_1, \ldots, \hat{o}_M$ that satisfy:
\begin{equation}
	\abs{\hat{o}_i - \Tr[\rho O_i]} \leq \epsilon \textrm{, for all } 1 \leq i \leq M
\end{equation}
with probability at least $1-\delta'$ over the randomness of the measurements of $\rho$, assuming that
 \begin{equation}
	N' \geq \O\left(\frac{\max_i \Tr[O_i^2]\log(M/\delta')}{\epsilon^2}\right).
\end{equation}
\end{lem}

\begin{lem}[Sample complexity of Pauli shadows (Theorem S1, Example 2 in \cite{huang2020predicting})]\label{lem:sampPauli}
Let $O_1, \ldots, O_M$ be a collection of $M$ Hermitian operators such that $\norm{O_i}_\infty \leq 1$, $\forall i$, and $\epsilon, \delta \in [0,1]$ be accuracy parameters. Pauli shadows using $N'$ copies of a quantum state $\rho$ allow to compute estimates $\hat{o}_1, \ldots, \hat{o}_M$ that satisfy:
\begin{equation}
	\abs{\hat{o}_i - \Tr[\rho O_i]} \leq \epsilon \textrm{, for all } 1 \leq i \leq M
\end{equation}
with probability at least $1-\delta'$ over the randomness of the measurements of $\rho$, assuming that
 \begin{equation}
	N' \geq \O\left(\frac{\max_i 4^{\textnormal{locality}(O_i)}\log(M/\delta')}{\epsilon^2}\right).
\end{equation}
\end{lem}

\medskip

\subsubsection{Deep measurements}\label{app:DeepMeasurements}

We start by analyzing the sample complexity of simulating the coherent training loss for $L$ parameter settings:

\begin{lem}[Simulation of the coherent training loss using deep measurements]\label{lem:simDeep}
Let $U$ be any $n$-qubit unitary and $V(\vec{\theta})$ a parametrized $n$-qubit unitary. Let $\epsilon,\delta>0$ and $N,L\in\mathbb{N}^*$. The global training cost $C_{\D(N)}(\vec{\theta})$ defined in Eq.~(\ref{eq:global-training}) can be estimated using Clifford shadows for an arbitrary sequence of parameter settings $\thv^{(1)}, \ldots, \thv^{(L)}$, up to additive error $\epsilon$ and with probability $1-\delta$ over the randomness of the shadowing procedure, using 
\begin{equation}
	\widetilde{\O} \left( \frac{N\log(NL/\delta)}{\epsilon^2} \right)
\end{equation}
calls to the unitary $U$.
\end{lem} 
\begin{proof}
For each training state $\ket{\psi^{(j)}}, 1 \leq j \leq N$, used to compute $C_{\D(N)}(\vec{\theta})$ call
\begin{equation}
O_{j,1}, \ldots, O_{j,L} = V(\thv^{(1)})\ket{\psi^{(j)}}\bra{\psi^{(j)}}V^\dagger(\thv^{(1)}),\ \ldots,\  V(\thv^{(L)})\ket{\psi^{(j)}}\bra{\psi^{(j)}}V^\dagger(\thv^{(L)}).
\end{equation}
By applying Lemma \ref{lem:sampClifford} for $\rho = U\ket{\psi^{(j)}}\bra{\psi^{(j)}}U^\dagger$, the $M=L$ observables above, and a probability of failure $\delta' = \delta/N$, we get that
\begin{equation}
	N' \in \O\left(\frac{\log(LN/\delta)}{\epsilon^2}\right)
\end{equation}
calls to $U$ are sufficient to guarantee that the term $\Tr[U\ket{\psi^{(j)}}\bra{\psi^{(j)}}U^\dagger\ O^{(j)}(\thv)]$ is estimated to additive error $\epsilon$ for all $L$ parameter settings, with probability at least $1-\delta/N$.\\
From the union bound, we have:
\begin{align*}
 P\left(\bigvee_{j=1}^N \textnormal{``estimation } j \textrm{ fails''}\right) &\leq \sum_{j=1}^N P(\textnormal{``estimation } j \textrm{ fails''})\\
 	&\leq \delta.
\end{align*}
which implies that all estimations succeed with probability at least $1-\delta$.
Therefore, 
\begin{equation}
    C_{\D(N)}(\vec{\theta}) = 1 - \frac{1}{N}\sum_{i=1}^{N} \Tr[U\ket{\psi^{(j)}}\bra{\psi^{(j)}}U^\dagger\ V(\thv)\ket{\psi^{(j)}}\bra{\psi^{(j)}}V^\dagger(\thv)]
\end{equation}
is computed successfully to additive error $\epsilon$ for all $L$ parameter settings with probability $1-\delta$, using the total claimed sample complexity.
\end{proof}

We can now move to the proof of Theorem \ref{thm:powerofdeep} in the main text, exposed formally in the following theorem.

\begin{thm}[Learning efficiently representable unitaries using deep measurements]\label{thm:powerofdeep(ap)}
Call an $n$-qubit unitary efficiently representable if it can be implemented using a quantum circuit of size $\text{poly}(n)$. Let $\epsilon>0$. For every polynomial circuit size, there exists a parametrized $n$-qubit unitary $V(\vec{\theta})$ that can be trained to achieve a Hilbert-Schmidt cost $C_{\rm HST}(\thv_{\rm opt})\leq \epsilon$  for any efficiently-representable $U$ of that size, with high probability over the randomness of the shadowing procedure and the choice of training states, using 
\begin{equation}
	\widetilde{\O} \left( \frac{\text{poly}(n)}{\epsilon^4} \right)
\end{equation}
calls to the unitary $U$.
\end{thm}
\begin{proof}

We first show how it is possible to construct an ansatz that captures all efficiently representable $n$-qubit unitaries of a given size. Assume that we are interested in unitaries that can be implemented using a quantum circuit of at most $n^c$ two-qubit gates with $c>0$. Then our ansatz is composed of $n^c$ applications of the following gadget $G$: a universal 2-qubit gate (see e.g., \cite{vatan2004optimal}) applied on the first two qubits, sandwiched by $2n$ discretely parametrized gates SWAP$^{b}$, $b\in\{0,1\}$, that can swap the first two qubits with any of the other qubits. This gadget allows one to apply an arbitrary 2-qubit gate to any pair of qubits in the circuit, and therefore, the entire circuit captures all unitaries that can be implemented using at most $n^c$ two-qubit gates.

We now construct the covering net for this ansatz with respect to the diamond distance $\norm{\mathcal{U} - \mathcal{V}}_{\diamond}$ (between the unitary channels $\mathcal{U},\mathcal{V}$ corresponding to the unitaries $U$ and $V$). We start from the covering net $\widetilde{\N}_{\widetilde{\epsilon}}$ of an arbitrary 2-qubit unitary with respect to the spectral norm $\norm{U - V}_\infty$, for $\widetilde{\epsilon} = \frac{\epsilon}{2n^c}$. From Lemma \ref{lem:covernet}, we can take a covering net of size $\abs{\widetilde{\N}_{\widetilde{\epsilon}}} \leq (12n^c/\epsilon)^{32}$. From here, we construct the set of unitaries corresponding to the gadget $G$ above simply by counting all possible choices of pairs of qubits this unitary can be applied to. This set, corresponding to one application of the gadget G, has size $\abs{\widetilde{\N}_{\widetilde{\epsilon}}}\binom{n}{2}$. Applying this gadget $n^c$ times then results in a set of size of size $\abs{\N_\epsilon}=\left(\abs{\widetilde{\N}_{\widetilde{\epsilon}}}\binom{n}{2}\right)^{n^c} \leq \left((12n^c/\epsilon)^{32} n^2\right)^{n^c}$. It follows that
\begin{equation}\label{eq:bound-covering-net}
\log(\abs{\N_\epsilon}) \leq \widetilde{\O}(n^c\log(n^c/\varepsilon)).
\end{equation}
We now proceed to show that this $\abs{\N_\epsilon}$ sized set of unitaries corresponding to $n^c$ applications of the gadget G is indeed an $\epsilon$-covering net of the ansatz with respect to the diamond distance. To do so, we first note that the unitaries in the set take the form $\widetilde{V}=U_{n_c}\widetilde{V}_{n^c}U_{n^c-1}\ldots U_1\widetilde{V}_1U_0$ where the $U_i$ are composed of SWAP gates and the $\widetilde{V}_i$ are 2-qubit unitaries, and they cover elements of the form $V(\thv)=U_{n_c}V_{n^c}(\thv)U_{n^c-1}\ldots U_1V_1(\thv)U_0$. Now consider such a unitary $V(\thv)$ and its corresponding $\widetilde{V}$ in the covering net obtained by replacing $V_1(\thv), \ldots, V_{n^c}(\thv)$ with the $\widetilde{\epsilon}$-close unitaries $\widetilde{V}_1, \ldots, \widetilde{V}_{n^c}$ in $\widetilde{\N}_{\widetilde{\epsilon}}$. We use the sub-additivity of the diamond distance and the relation between diamond distance and spectral norm distance (see Eq.~(C8) in \cite{caro2021generalization}) to show:
\begin{equation}
\norm{\widetilde{\mathcal{V}} - \mathcal{V}(\thv)}_{\diamond} \leq \sum_{i=1}^{n^c} \norm{\widetilde{\mathcal{V}}_i - \mathcal{V}_i(\thv)}_{\diamond} + \sum_{i=0}^{n^c} \norm{\mathcal{U}_i - \mathcal{U}_i}_{\diamond} \leq 2 \sum_{i=1}^{n^c} \norm{\widetilde{V}_i - V_i(\thv)}_{\infty} \leq \epsilon.
\end{equation}
From here we consider the parameter settings $\thv^{(1)}, \ldots, \thv^{(L)}$ for $L=\abs{\N_\epsilon}$ such that $V(\thv^{(1)}), \ldots, V(\thv^{(L)})$ correspond to the elements of the covering net $\N_\epsilon$. Since the target unitary $U$ is guaranteed to correspond to some assignment $\thv_U$, then by definition of the covering net, there exist $1\leq i \leq L$ such that $\norm{\mathcal{V}(\thv^{(i)}) - \mathcal{U}}_{\diamond}\leq\epsilon$. We make use of the definition of the diamond norm and its relation to the fidelity between pure states for unitary channels \cite{wilde2011classical} to show:
\begin{equation}
    \begin{aligned}
        ||\mathcal{V}(\thv^{(i)}) - \mathcal{U} ||_{\diamond} &:= \max_\psi  ||V(\thv^{(i)}) |\psi \rangle \langle \psi | V^\dagger(\thv^{(i)}) - U  |\psi \rangle \langle \psi | U^\dagger ||_{1} \\ &= \max_\psi \left( 1  - | \langle \psi | V^\dagger(\thv^{(i)}) U  |\psi \rangle |^2 \right) \\ &\geq
        1  - \frac{1}{N} \sum_{j=1}^N | \langle \psi^{(j)} | V^\dagger(\thv^{(i)}) U  | \psi^{(j)}  \rangle |^2 \\
        &:=  C_{\D(N)}(\thv^{(i)}) \, .
    \end{aligned}
    \end{equation}
And therefore, there exists a parameter setting $\thv^{(i)}$ in the covering net that achieves a training error $C_{\D(N)}(\thv^{(i)}) \leq \epsilon$. We also make use of the generalization bounds for this ansatz \cite{caro2021generalization,caro2022outofdistribution}.
With high probability,
\begin{equation}
        C_{\rm HST}(\thv^{(i)}) \leq 2\frac{d+1}{d}\epsilon + \mathcal{O} \left( \sqrt{\frac{T \log (T)}{N}} \right)
\end{equation}
for $T=2n^{c+1}+15n^c$ its number of parameters (given that a universal 2-qubit gate has 15 parameters \cite{vatan2004optimal}, and that we use $2n$ parametrized SWAP gates per gadget $G$). This means that we can guarantee $C_{\rm HST}(\thv^{(i)}) \leq \O(\epsilon)$ with high probability using 
\begin{equation}\label{eq:bound-N}
N \in \widetilde{\O}\left(\frac{n^{c+1}}{\epsilon^2}\right)
\end{equation}
training states. Then, by applying Lemma \ref{lem:simDeep} for $\thv^{(1)}, \ldots, \thv^{(L)}$ ($L=\abs{\N_\epsilon}$) defined above, and using the bounds in Eq.~(\ref{eq:bound-covering-net}) and (\ref{eq:bound-N}), we get a total sample complexity in 
\begin{equation}
\widetilde{\O} \left( \frac{n^{2c+1}}{\epsilon^4} \right).
\end{equation}
\end{proof}

\vspace{-1em}

\subsubsection{Shallow measurements}\label{app:ShallowMeasurements}

We also prove analogous results for the case of shallow measurements.

\begin{lem}[Simulation of the coherent training loss using shallow measurements]\label{lem:simShallow}
Let $U$ be any $n$-qubit unitary and $V(\vec{\theta})$ a parametrized $n$-qubit unitary that uses $T$ trainable gates and is low-entangling, i.e., it transforms $1$-local operators into $\O(\log(n))$-local operators. Let $\epsilon,\delta>0$ and $N,L\in\mathbb{N}^*$. The local training cost $C_N^\ell(\vec{\theta})$ defined in Eq.~(\ref{eq:local-training}) can be estimated using Pauli shadows for an arbitrary sequence of parameter settings $\thv^{(1)}, \ldots, \thv^{(L)}$, up to additive error $\epsilon$ and with probability $1-\delta$ over the randomness of the shadowing procedure, using 
\begin{equation}
	\widetilde{\O} \left( \frac{N\log(NLn/\delta)}{\epsilon^2} \right)
\end{equation}
calls to the unitary $U$.
\end{lem} 
\begin{proof}
The proof only differs for that of Lemma \ref{lem:simDeep} in the definition of the observables:
\begin{equation}
O_{i,j,l} =  V(\thv) \left(\ket{\psi_i^{(j)}} \bra{\psi_i^{(j)} }\otimes \mathds{1}_{\bar{i}}\right) V^\dagger(\thv),\quad 1\leq i \leq n, 1 \leq j \leq N, 1\leq l \leq L,
\end{equation}
where we have now $M=nL$ observables per training state, and the application of Lemma \ref{lem:sampPauli} instead of Lemma \ref{lem:sampClifford}.
\end{proof}

The following theorem is a formal exposition of Theorem \ref{thm:powerofshallow} in the main text.

\begin{thm}[Learning low-entangling unitaries using shallow measurements]\label{thm:powerofshallow(ap)}
Call an $n$-qubit unitary low-entangling if it transforms $1$-local operators into $\O(\log(n))$-local operators. Let $\epsilon>0$. There exists a parametrized $n$-qubit unitary $V(\vec{\theta})$ that can be trained to achieve a Hilbert-Schmidt cost $C_{\rm HST}(\thv_{\rm opt})\leq \epsilon$ for any low-entangling $U$ with high probability over the randomness of the shadowing procedure and the choice of training states, using 
\begin{equation}
	\widetilde{\O} \left( \frac{\text{poly}(n)}{\epsilon^4} \right)
\end{equation}
calls to the unitary $U$.
\end{thm}
\begin{proof}
We apply the same reasoning as in Theorem \ref{thm:powerofdeep(ap)} but where we keep only low-entangling unitaries in the covering net, which results, using Lemma \ref{lem:simShallow}, in a similar sample complexity.
\end{proof}

\subsubsection{Approximate locality}\label{ap:approxlocal}

In this section we argue that our proposed shallow measurement incoherent learning algorithm works not only for unitaries that are low entangling in the sense of transforming $1$-local operators into $\O(\log(n))$-local operators but also ones that transform $1$-local operators into \textit{approximately} $\O(\log(n)$)-local operators. That is, we pin down the intuition, supported by our numerics in Fig.~\ref{fig:Simulation_plot}, that terms in the observables $O_i$ that are highly non-local but have a small weight do not contribute substantially to the scaling of our algorithm. 

\medskip
To see this, we first formally define what we mean by approximately $k$-local.
\begin{defn}[Approximate locality- $(\alpha, k)$-locality]
Let $O_i = \tilde{O}_i(k) +\chi_i(k)$ where $\max_i{\textnormal{locality}\left(O_i(k)\right)} = k$ and $\min_i{\textnormal{locality}\left(\chi_i(k)\right)} = k+1$ . Then we say an operator $O_i$ is $(\alpha, k)$-local if $||  \chi_i(k) ||_{\infty} = \alpha$.
\end{defn}

Equipped with this definition, we can rewrite Theorem~\ref{thm:powerofshallow(ap)} as follows.

\begin{thm}[Learning low-entangling unitaries using shallow measurements: $(\alpha, k)$-local version]\label{thm:powerofshallow-alphalocal}
Call an $n$-qubit unitary low-entangling $n$-qubit unitary if it transforms $1$-local operators into  $(\alpha, k)$-local operators with $k \in \O(\log(n))$. Let $\epsilon>0$. There exists a parametrized $n$-qubit unitary $V(\vec{\theta})$ that can be trained to achieve a Hilbert-Schmidt cost $C_{\rm HST}(\thv_{\rm opt})\leq \epsilon + \alpha$ for any low-entangling $U$ with high probability over the randomness of the shadowing procedure and the choice of training states, using 
\begin{equation}
	\widetilde{\O} \left( \frac{\text{poly}(n)}{(\epsilon + \alpha)^4} \right)
\end{equation}
calls to the unitary $U$.
\end{thm}

\proof{The core observation is that 
\begin{equation}
    \begin{aligned}\label{eq:alphalocalitybound}
        \abs{\hat{o}_i - \Tr[\rho O_i]} = \abs{\hat{o}_i - \Tr[\rho \tilde{O}_i(k)] -  \Tr[\rho  \chi_i(k) ] } \leq \abs{\hat{o}_i - \Tr[\rho \tilde{O}_i(k)]} + \abs{ \Tr[\rho  \chi_i(k) ] } \leq  \abs{\hat{o}_i - \Tr[\rho \tilde{O}_i(k)]} + \alpha \, ,
    \end{aligned}
\end{equation}
where we use the triangle inequality and $|\Tr[\rho  \chi_i(k) ] | \leq ||  \chi_i(k)  ||_{\infty} $. The remainder of the proof is identical to that of Theorem~\ref{thm:powerofshallow(ap)}. 

That indeed $(\alpha,k)$-locality, and not standard locality (i.e. $\alpha = 0$ locality), determines the ultimate scaling of our algorithm (as claimed in Theorem~\ref{thm:powerofshallow-alphalocal}) is seen in the numerics shown in Fig.~\ref{fig:Simulation_plot} in the main text. There we study the performance of our learning algorithm as a function of shadow size for different evolution times of the target unitary ($\Delta t$) and different target unitary depths. If standard locality were what mattered, then we would expect the complexity of the algorithm to depend directly on the depth of the guess/target unitary. This is because the width of the light-cone of the ansatz grows linearly with depth of the guess unitary, and so the locality of the parameterized observables (Eq.~\eqref{eq:paramobservables}), similarly increase linearly with depth. However, in Fig.~\ref{fig:Simulation_plot} the learning performance hardly changes as we increase from a 1-layer target to a 2-layer target. Instead the performance depends on $\Delta t$. This makes sense because $\Delta t$ determines how entangling the guess/target unitary is and hence the effective light cone of the guess/target unitary once small tails have been disregarded. In other words, $\Delta t$ determines the $(\alpha,k)$-locality of the parameterized observables and thus the complexity of the algorithm.

}

\begin{figure*}[t!]
\centering
\includegraphics[width =0.95\columnwidth]{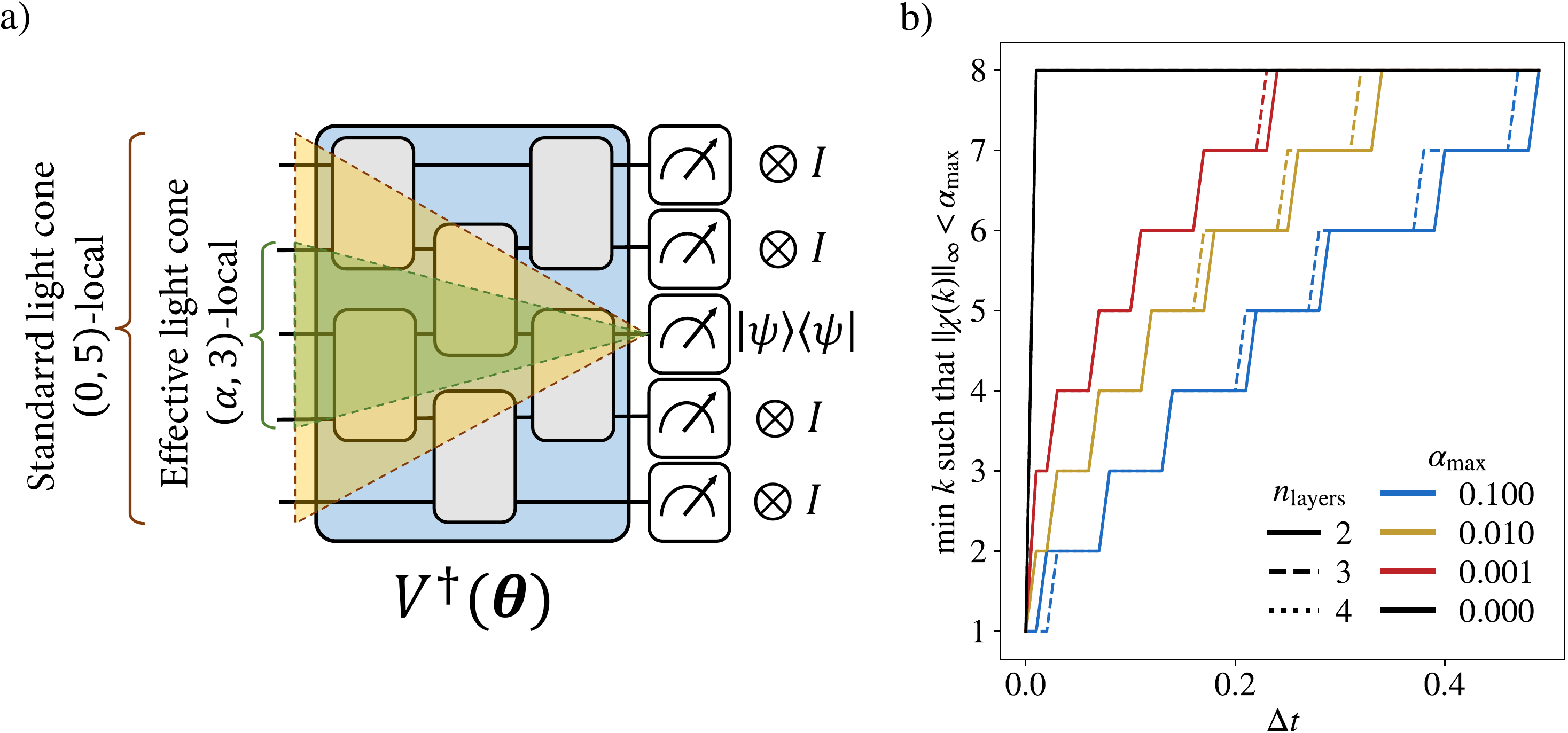}
\vspace{-4mm}
\caption{\textbf{Standard and approximate locality.} a) Cartoon representation of the difference between standard and approximate locality. b) Here we compare the growth of standard locality, i.e. $(0, k)$-locality, to approximate locality, i.e. $(\alpha, k)$-locality. We focus on the locality of the operator $V(\Delta t)^{\dagger}(I_{\frac{n}{2}-1}\otimes|\psi\rangle\langle\psi|\otimes I_{\frac{n}{2}})V(\Delta t)$, where $V(\Delta t) = U(\Delta t /  n_{\text layers})^{n_{\text layers}}$ and $U(\Delta t)$ is the first order Trotter operator for the 8 qubit 1D Heisenberg Hamiltonian with open boundary conditions, and $|\psi\rangle$ is a single qubit Haar random state. While standard locality only depends on the number of layers $n_{\rm layers}$ and not the evolution time $\Delta t$, the approximate locality is strongly dependent on the overall evolution time, and not dependent on the total number of layers. The upper-limit of $k$ is the number of qubits in the system, which is 8 in our example.
}
\label{fig:k_epsilon}
\end{figure*}

\subsection{Remarks on the computational complexity}\label{app:ComputComplexIncoherent}

So far, we've only discussed the sample complexity of our incoherent learning protocols. In this section, we point out the caveats that appear when looking at their computational complexity.

The most obvious caveat of both protocols has to do with the exhaustive search used in both Theorem \ref{thm:powerofdeep(ap)} and Theorem \ref{thm:powerofshallow(ap)}. While the scaling of the sample complexity is logarithmic in terms of the number $L$ of parameter settings considered, the scaling remains linear in $L$ for the computational complexity. Since $L$ is exponential in $n$, this becomes quickly intractable.

However, we would like to consider here other computational aspects of our protocols, when used to train parametrized circuits in a standard fashion (e.g., using gradient-based optimization). As mentioned in the main text, the deep measurement setting suffers from several pitfalls. While Lemma \ref{lem:sampClifford} guarantees that the expectation values
\begin{equation}\label{eq:term-global-loss}
    \Tr[U\ket{\psi^{(j)}}\bra{\psi^{(j)}}U^\dagger\ V(\thv)\ket{\psi^{(j)}}\bra{\psi^{(j)}}V^\dagger(\thv)]
\end{equation}
can be estimated efficiently in terms of sample complexity, the estimation algorithm in Ref.~\cite{huang2020predicting} still needs to compute the observables $O^{(j)}(\thv) = V(\thv)\ket{\psi^{(j)}}\bra{\psi^{(j)}}V^\dagger(\thv)$ classically in order to compute their inner product with the classical shadows $\rho_m^{{(j)}}$ in Eq.~(\ref{eq:shadowClifford}).\footnote{Note that the operators $O^{(j)}(\thv) = V(\thv)\ket{\psi^{(j)}}\bra{\psi^{(j)}}V^\dagger(\thv)$ can be viewed as observables but also as quantum states. In this paragraph, we make use of both of these views, whenever they are more convenient.} For a general $V(\thv)$, these observables can represent arbitrary rank-$1$ density matrices, and therefore are neither easy to represent nor to compute classically. One could imagine here making use of Clifford shadows again to prepare states $O^{(j)}(\thv)$ on a quantum computer, collect their shadows, and estimate their inner products with $\rho_m^{{(j)}}$. However, since $\rho_m^{{(j)}}$ is always a high-rank observable ($\text{Tr}[(\rho_m^{{(j)}})^2]=2^{n+1}-1$), then Clifford shadows have no guarantees of estimating this inner product efficiently. More generally, any protocol that would only use measurements outcomes of the density matrices $O^{(j)}(\thv)$ without knowledge of how these states were generated would be at risk of facing the $\Omega(\sqrt{2^n}/\epsilon)$ lower bound in sample complexity of incoherently estimating the overlap of two unknown input quantum states to $\epsilon$ precision (see Theorem 5 in Ref.~\cite{anshu2022distributed}). Similarly, if one would attempt preparing the classical shadows $\rho_m^{{(j)}}$ on a quantum computer, evolve them under $V^\dagger(\thv)$, and measure the expectation value of $\ket{\psi^{(j)}}\bra{\psi^{(j)}}$, i.e.,
\begin{equation}
    \Tr[V^\dagger(\thv)\rho_m^{{(j)}}\ V(\thv)\ket{\psi^{(j)}}\bra{\psi^{(j)}}],
\end{equation}
which is equivalent to the expression in Eq.~(\ref{eq:term-global-loss}), one would face the problem that the classical shadows $\rho_m^{{(j)}}$ are completely unphysical states: while they have unit trace, they are not PSD, and have trace norm $||\rho_m^{{(j)}}||_1 = 2^{n+1}-1$.

As for our shallow measurement protocol, aside form the exhaustive search use in Theorem \ref{thm:powerofshallow(ap)}, the simulation of the coherent training loss in Lemma \ref{lem:simShallow} is computationally efficient. Indeed, as one can visualize in Fig.~\ref{fig:lightcone}, the observables $O^{(j)}_i(\thv) = V(\thv) \left(\ket{\psi_i^{(j)}} \bra{\psi_i^{(j)} }\otimes \mathds{1}_{\bar{i}}\right) V^\dagger(\thv)$ are $\O(\log(n))$-local, and the classical shadows in Eq.~(\ref{eq:shadowPauli}) have a tensor-product form. Then we can simply classically simulate the circuit $V(\thv)$ on a $\O(\log(n))$-qubit region (i.e., a $\O(\poly(n))$ Hilbert space), and efficiently compute the expectation values of $O^{(j)}_i(\thv)$. Moreover, since the entire training loss $C_{\rm N}^{\ell}$ (Eq.~(\ref{eq:local-training})) can therefore be computed classically, it can also be trained classically. We note as well that the training loss that we use in this protocol only contains local terms, therefore it is less prone to barren plateaus \cite{cerezo2020cost}.\break

\section{Shallow measurements fail in learning linear depth circuits}\label{app:lower-bounds-shallow}

\begin{figure*}
\centering
\includegraphics[width =0.9\columnwidth]{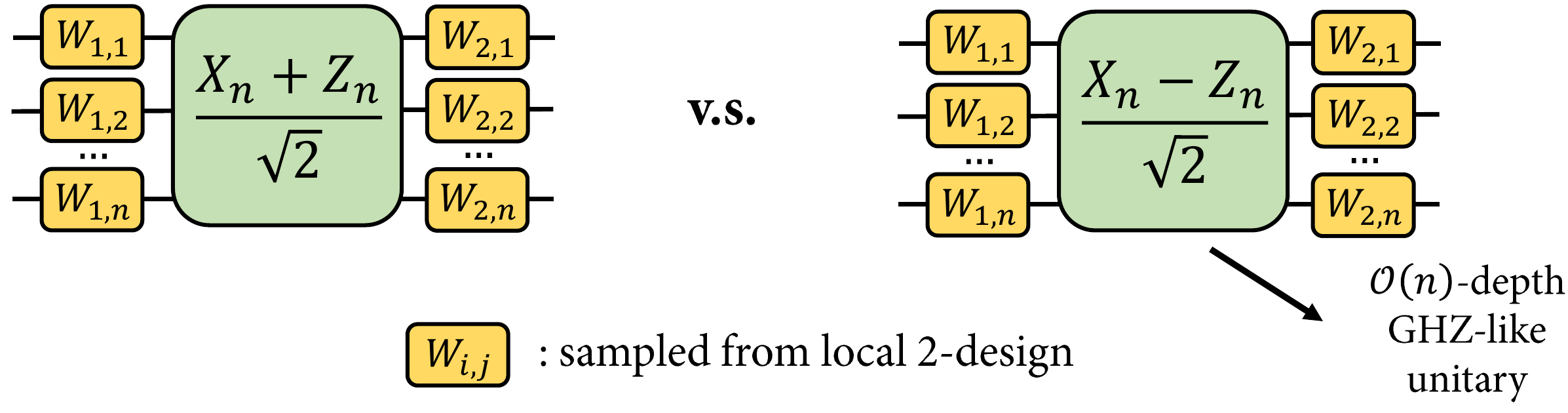}
\vspace{-4mm}
\caption{\textbf{Learning task used to show the limitations of shallow measurements.} We show the hardness of learning polynomial-depth unitaries using shallow measurements by reduction of the problem of distinguishing between the unitaries $U_+ = (X^{\otimes n} + Z^{\otimes n})/\sqrt{2}$ and $U_- = (X^{\otimes n} - Z^{\otimes n})/\sqrt{2}$ (for $n$ odd). We first consider the warm-up case where these target unitaries are taken as is (Lemma \ref{lem:pauli-compilation-lower-bound}), followed by the harder case where they are sandwiched with products of unitaries sampled from a local 2-design (Theorem \ref{thm:pauli-compilation-lower-bound}). We call $U_+$ and $U_-$ ``GHZ-like'' unitaries because they prepare orthogonal GHZ states $(\ket{0}^{\otimes n} \pm \ket{1}^{\otimes n}))/\sqrt{2}$ when they are applied on a $\ket{0}^{\otimes n}$ state.}
\label{fig:distinguisging-task}
\end{figure*}

\subsection{Warm-up example}

We start by studying the simpler case where the shallow measurements considered are limited to be random Pauli measurements, as those used in Pauli shadows.

\begin{lem}\label{lem:pauli-compilation-lower-bound}
Consider the product-POVM $F = \frac{1}{3^n}\{\kb{0}, \kb{1}, \kb{+}, \kb{-}, \kb{i}, \kb{-i}\}^{\otimes n}$ which corresponds to measuring in the eigenbasis of a random $n$-qubit Pauli observable in $\{X,Y,Z\}^{\otimes n}$, and a batch $\D$ of $T$ quantum states sampled uniformly from $\{\0, \1, \+, \-, \i, \mi\}^{\otimes n}$. Suppose there exists an algorithm that, for any unitary $U$ that can be implemented in $\O(n)$ depth, returns (the description of) a unitary $V$ such that $1-\frac{1}{4^n}\abs{\Tr[UV^\dagger]}^2 \leq 1/4$ with high probability, using, for each quantum state $\Psiin$ in $\D$, the measurement outcomes of the POVM $F$ on $M$ copies of $U\Psiin$. Then, necessarily, the total number of measurements that any such algorithm must use is
\begin{equation}
    MT \geq 2^{\Omega(n)}.
\end{equation}
\end{lem}

\begin{proof}
We prove this result by reduction of the task of distinguishing between the unitaries (see Figs.~\ref{fig:distinguisging-task} and \ref{fig:GHZ-unitaries}):
\begin{equation}\label{eq:GHZ-unitaries}
\begin{cases}
U_+ = (X^{\otimes n} + Z^{\otimes n})/\sqrt{2}\quad \text{and} \quad U_- = (X^{\otimes n} - Z^{\otimes n})/\sqrt{2} \quad\quad \text{for }n\text{ odd}\\
U_+ = (X^{\otimes n} + iZ^{\otimes n})/\sqrt{2}\quad \text{and} \quad U_- = (X^{\otimes n} - iZ^{\otimes n})/\sqrt{2}\quad\quad \text{for }n\text{ even}\\
\end{cases}
\end{equation}
which we show requires $MT \geq 2^{\Omega(n)}$ measurements.

\paragraph{Reduction}
Assume that an algorithm can achieve $\frac{1}{4^n}\abs{\Tr[U_+V^\dagger]}^2 \geq 3/4$, when $U_+$ is the target unitary. We first note that
\begin{equation}
    \frac{1}{4^n}\abs{\Tr[U_+U_-^\dagger]}^2 = 0.
\end{equation}
Therefore, we can decompose $V$ in a $\{U_+, U_-, \ldots\}$ basis as
\begin{equation}
    V = \frac{1}{2^n}\Tr[VU_+^\dagger] U_+ + \frac{1}{2^n}\Tr[VU_-^\dagger] U_- + \ldots
\end{equation}
and deduce (from Parseval's identity) that
\begin{equation}
    1 = \frac{1}{4^n}\abs{\Tr[VV^\dagger]}^2 = \left(\frac{1}{4^n}\abs{\Tr[U_+V^\dagger]}^2\right)^2 + \left(\frac{1}{4^n}\abs{\Tr[U_-V^\dagger]}^2\right)^2 + c
\end{equation}
where $c \geq 0$. From here, we have:
\begin{align}
    \frac{1}{4^n}\abs{\Tr[U_-V^\dagger]}^2 & \leq \sqrt{1 - \left(\frac{1}{4^n}\abs{\Tr[U_+V^\dagger]}^2\right)^2}\\
    &\leq \sqrt{1 - \left(\frac{3}{4}\right)^2} < 0.67.
\end{align}
We find $\frac{1}{4^n}\abs{\Tr[U_-V^\dagger]}^2 < \frac{1}{4^n}\abs{\Tr[U_+V^\dagger]}^2$ when $U_+$ is the target unitary, and through a similar argument $\frac{1}{4^n}\abs{\Tr[U_+V^\dagger]}^2 < \frac{1}{4^n}\abs{\Tr[U_-V^\dagger]}^2$ when $U_-$ is the target unitary. Therefore, this algorithm can be used as a subroutine to distinguish between $U_+$ and $U_-$, simply by learning $V$, computing $\frac{1}{4^n}\abs{\Tr[U_+V^\dagger]}^2$ and $\frac{1}{4^n}\abs{\Tr[U_-V^\dagger]}^2$ and outputting the unitary with the largest overlap.\\

\begin{figure*}
\centering
\includegraphics[width =0.8\columnwidth]{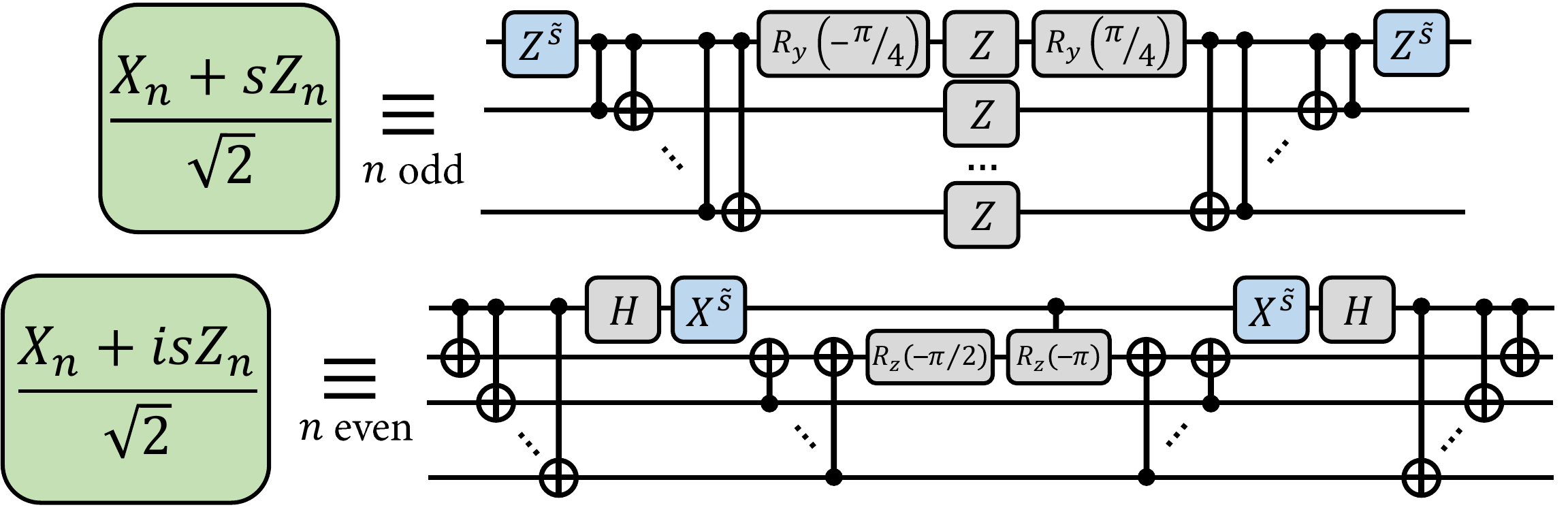}
\vspace{-2mm}
\caption{\textbf{Circuit decomposition of the GHZ-like unitaries.} We give a $\O(n)$-depth circuit decomposition of the unitaries $U_+$ and $U_-$ defined in Eq.~(\ref{eq:GHZ-unitaries}). We use the notation $s=\pm 1$ and $\widetilde{s}=\frac{s+1}{2}=0/1$. The resulting unitaries are equivalent up to a global phase $s$.}
\label{fig:GHZ-unitaries}
\vspace{-2mm}
\end{figure*}

\paragraph{Hardness of distinguishing $U_+$ and $U_-$}
For $U_\pm \in \{U_+, U_-\}$, consider the joint probability distribution 
\begin{equation}
    p_\pm(\psi_1, \ldots, \psi_T, \phi_1^1, \ldots, \phi_T^M) = p_\pm(\psi_{1:T}, \phi_{1:T}^{1:M})
\end{equation}
of sampling the input states
\begin{equation}
    \ket{\psi_1}, ..., \ket{\psi_T} \in \left(\{\0, \1, \+, \-, \i, \mi\}^{\otimes n}\right)^T
\end{equation}
when generating the batch $\D$, and of measuring the projections
\begin{equation}
    \kb{\phi_1^1}, \ldots, \kb{\phi_T^M} \in \left(\{\kb{0}, \kb{1}, \kb{+}, \kb{-}, \kb{i}, \kb{-i}\}^{\otimes n}\right)^{MT}
\end{equation}
respectively on the $M$-copy blocks of these states, after being all evolved under $U_\pm$.\\
LeCam's method (see, e.g., Lemma 1 in \cite{yu1997assouad}) tells us that the probability of success of any algorithm that attempts to distinguish between $U_+$ and $U_-$ with the type of access defined in the statement of this Lemma is upper bounded by:
\begin{equation}
    \textrm{P[success]} \leq \frac{1}{2} + \frac{1}{2}\TV(p_+,p_-)
\end{equation}
where
\begin{equation}
    \TV(p_+,p_-) = \frac{1}{2} \sum_{\psi_{1:T}, \phi_{1:T}^{1:M}} \abs{p_+(\psi_{1:T}, \phi_{1:T}^{1:M}) - p_-(\psi_{1:T}, \phi_{1:T}^{1:M})}.
\end{equation}
Therefore, to show that $MT \geq 2^{\Omega(n)}$ measurements are necessary to distinguish the two unitaries with high probability, our goal is to show that this total variation distance between the two joint probability distributions is of the order $MT/2^{\O(n)}$.\\

We establish this result using a very handy property from probability theory. Namely, if we take $p_i, q_i$ to be probability distributions on a finite space $\Omega_i, i=1,\ldots,N$, then, for the product distributions $p = \prod_{i=1}^N p_i$ and $q = \prod_{i=1}^N q_i$ on the finite space $\Omega = \prod_{i=1}^N \Omega_i$, we have \cite{tvproduct}:
\begin{equation}\label{eq:TV-product}
    \TV(p,q) \leq \sum_{i=1}^N \TV(p_i,q_i).
\end{equation}

From here, we note that all of the $T$ $M$-copy blocks of measurements are i.i.d., and that inside each of these blocks, the $M$ measurements are also i.i.d., conditioned on the choice of $\ket{\psi_i}$. Therefore, we can write
\begin{align}
    \TV(p_+, p_-) &= \frac{1}{2} \sum_{\psi_{1:T}, \phi_{1:T}^{1:M}} \abs{p_+(\psi_{1:T}, \phi_{1:T}^{1:M}) - p_-(\psi_{1:T}, \phi_{1:T}^{1:M})}\nonumber\\
    &\leq \frac{T}{2} \sum_{\psi_{1}, \phi_{1}^{1:M}} \abs{p_+(\psi_{1}, \phi_{1}^{1:M}) - p_-(\psi_{1}, \phi_{1}^{1:M})}\nonumber\\
    &= \frac{T}{2} \sum_{\psi_{1}, \phi_{1}^{1:M}} p(\psi_1) \abs{p_+(\phi_{1}^{1:M} | \psi_{1}) - p_-(\phi_{1}^{1:M} | \psi_{1})}\nonumber\\
    &\leq \frac{MT}{2} \sum_{\psi_{1}, \phi_{1}} p(\psi_1) \abs{p_+(\phi_{1}^{1}| \psi_{1}) - p_-(\phi_{1}^{1} | \psi_{1})},\label{eq:tvd-ineq-pauli}
\end{align}
which means we only need to focus on the total variation distance of a single measurement. Here, we made use of Eq.~(\ref{eq:TV-product}) in the second and fourth lines, and of the decomposition $p_\pm(\psi_{1}, \phi_{1}^{1:M})=p(\psi_1)p_\pm(\phi_{1}^{1:M} | \psi_{1})$ in the third line.

For $n$ odd, the total variation distance of a single measurement is given by
\begin{align}\label{eq:deriv-single-TV}
    \TV(p_+^1, p_-^1) &= \frac{1}{4} \sum_{\psi, \phi} \frac{p(\psi)}{3^n} \abs{\Tr[(X_n\!+\!Z_n)\kb{\psi}(X_n\!+\!Z_n)\kb{\phi}] - \Tr[(X_n\!-\!Z_n)\kb{\psi}(X_n\!-\!Z_n)\kb{\phi}]}\nonumber\\
    &= \frac{1}{2} \sum_{\psi, \phi} \frac{p(\psi)}{3^n} \abs{\Tr[X_n\kb{\psi}Z_n\kb{\phi}] + \Tr[Z_n\kb{\psi}X_n\kb{\phi}]}\nonumber\\
    &= \sum_{\psi, \phi} \frac{p(\psi)}{3^n} \abs{\Re[\bra{\phi}X_n\kb{\psi}Z_n\ket{\phi}]},
\end{align}
where we have abbreviated $X^{\otimes n}$ and $Z^{\otimes n}$ by $X_n$ and $Z_n$, and have dropped the sub- and super- scripts of $\ket{\psi_{1}}$ and $\ket{\phi_{1}^{1}}$.\break Similarly, for $n$ even, we have
\begin{equation}
    \TV(p_+^1, p_-^1) = \sum_{\psi, \phi} \frac{p(\psi)}{3^n} \abs{\Im[\bra{\phi}X_n\kb{\psi}Z_n\ket{\phi}]}.
\end{equation}
For all $n$, we then have
\begin{equation}
    \TV(p_+^1, p_-^1) \leq \sum_{\psi, \phi} \frac{p(\psi)}{3^n} \abs{\bra{\phi}X_n\kb{\psi}Z_n\ket{\phi}}.
\end{equation}
To show that this quantity is exponentially small in $n$, we exploit the tensor product structure of $\ket{\psi}$ and $\ket{\phi}$ as well as a property of the stabilizer states $\S=\{\0, \1, \+, \-, \i, \mi\}$ that is easy to check (numerically):
\begin{equation}
    \E_{\ket{x}, \ket{y} \in \S}[\abs{\bra{x}X\kb{y}Z\ket{x}}] = \frac{1}{36}\sum_{\ket{x}, \ket{y} \in \S} \abs{\bra{x}X\kb{y}Z\ket{x}} = \frac{7}{18}.
\end{equation}
This gives us 
\begin{align}
    \TV(p_+^1, p_-^1) &\leq \sum_{\psi, \phi} \frac{p(\psi)}{3^n} \abs{\bra{\phi}X_n\kb{\psi}Z_n\ket{\phi}}\nonumber\\
    &= \sum_{\psi, \phi} \frac{1}{3^n}\prod_{i=1}^{n} p(\psi^{(i)}) \abs{\bra{\phi^{(i)}}X\kb{\psi^{(i)}}Z\ket{\phi^{(i)}}}\nonumber\\
    &= \sum_{\psi, \phi} \frac{1}{18^n}\prod_{i=1}^{n} \abs{\bra{\phi^{(i)}}X\kb{\psi^{(i)}}Z\ket{\phi^{(i)}}}\nonumber\\
    &= \frac{1}{18^n}\prod_{i=1}^{n} \sum_{\psi^{(i)}, \phi^{(i)} \in \S} \abs{\bra{\phi^{(i)}}X\kb{\psi^{(i)}}Z\ket{\phi^{(i)}}}\nonumber\\
    &= \frac{36^n}{18^n}\left(\frac{7}{18}\right)^n = \left(\frac{14}{18}\right)^n.
\end{align}
Combining this last inequality with Eq.~(\ref{eq:tvd-ineq-pauli}), we get:
\begin{equation}
    \TV(p_+, p_-) \leq MT \left(\frac{14}{18}\right)^n < MT/2^{0.36n}
\end{equation}
and therefore $MT$ must be scaling as $2^{\Omega(n)}$ for the algorithm to have high probability of success.
\end{proof}

Note, however, that when we step away from randomly chosen input states and measurement bases, distinguishing efficiently between $U_+$ and $U_-$ using product input states and product measurements becomes possible. In the case where $n$ is odd for instance, it is sufficient to consider the input state $\ket{\psi} = \ket{+}^{\otimes n}$, and to measure $U\ket{\psi}$ in the computational basis. One can then check that the states $U_+\ket{\psi}$ and $U_-\ket{\psi}$ have completely disjoint supports, which allows to distinguish between the two unitaries using only 1 measurement. A similar observation can be made for the case when $n$ is even, by taking for instance $\ket{\psi} = \ket{+}^{\otimes n-1}\otimes\ket{i}$. This justifies why we look into a more general case (and a harder learning task) in the next section.

\subsection{General case}

We now move to the general setting of arbitrary product measurements, that are moreover allowed to be adaptively chosen (i.e., the next input state and measurement basis can be chosen as a function of previous measurement outcomes).

\begin{thm}\label{thm:pauli-compilation-lower-bound}
Call $F_1, \ldots, F_T$ an arbitrary sequence of $n$-qubit product-POVMs, i.e., $F_i = \{w_{i,j}2^n\kb{\phi_{i,j}}\}_j$ where $\ket{\phi_{i,j}} = \ket{\phi_{i,j}^{(1)}} \otimes \ldots \otimes \ket{\phi_{i,j}^{(n)}}$ and $\sum_{j} w_{i,j} = 1$, and $\ket{\psi_1}, \ldots, \ket{\psi_T}$ an arbitrary sequence of product $n$-qubit quantum states, i.e., $\ket{\psi_i} = \ket{\psi_i^{(1)}} \otimes \ldots \otimes \ket{\psi_i^{(n)}}$, that can both be chosen adaptively (i.e., a choice of quantum state $\ket{\psi_i}$ and POVM $F_i$ can depend on the previous measurement outcomes resulting from the choices $\ket{\psi_1}, \ldots, \ket{\psi_{i-1}}$ and $F_1, \ldots, F_{i-1}$). Suppose there exists an algorithm that, for any unitary $U$ that can be implemented in $\mathcal{O}(n)$ depth, returns (the description of) a unitary $V$ such that $1-\frac{1}{4^n}\abs{\Tr[UV^\dagger]}^2 \leq 1/4$ with high probability, using the measurement outcomes of $F_1, \ldots, F_T$ on $U\ket{\psi_1}, \ldots, U\ket{\psi_T}$ respectively. Then, necessarily, the total number of measurements any such algorithm must use is
\begin{equation}
    T \geq 2^{\Omega(n)}.
\end{equation}
\end{thm}

\begin{proof}
Consider the ``GHZ-like'' unitaries (see Figs.~\ref{fig:distinguisging-task} and \ref{fig:GHZ-unitaries}):
\begin{equation}\label{eq:GHZ-unitaries2}
\begin{cases}
U_+ = (X^{\otimes n} + Z^{\otimes n})/\sqrt{2}\quad \text{and} \quad U_- = (X^{\otimes n} - Z^{\otimes n})/\sqrt{2} \quad\quad \text{for }n\text{ odd}\\
U_+ = (X^{\otimes n} + iZ^{\otimes n})/\sqrt{2}\quad \text{and} \quad U_- = (X^{\otimes n} - iZ^{\otimes n})/\sqrt{2}\quad\quad \text{for }n\text{ even}\\
\end{cases}
\end{equation}

We prove the result by reduction of the task of distinguishing between the unitaries
\begin{equation}
\begin{cases}
\hfil W_2U_+W_1\quad &\text{and} \quad \hfil W_2U_-W_1\\
\text{for }U_+ \text{ and }U_- \text{ given by Eq.~(\ref{eq:GHZ-unitaries2})}\quad &\text{and} \quad W_1 = \bigotimes_{i=1}^n W_1^{(i)}, W_2 = \bigotimes_{i=1}^n W_2^{(i)}
\end{cases}
\end{equation}
for some choice of single-qubit unitaries $W_{1/2}^{(1)}, \ldots, W_{1/2}^{(n)}$ taken from the single-qubit Clifford group $\mathcal{C}_1$, i.e., $(W_1,W_2) \in \W = \mathcal{C}_1^{\otimes n} \times \mathcal{C}_1^{\otimes n}$, which are revealed to the learner \emph{after} the $T$ measurements.\\

\paragraph{Reduction}
The reduction is similar to that of our previous Lemma, by noting that, for any unitaries $W_1, W_2$, we still have
\begin{equation}
    \frac{1}{4^n}\abs{\Tr[W_2U_+W_1 W_1^\dagger U_-^\dagger W_2^\dagger]}^2 = \frac{1}{4^n}\abs{\Tr[U_+U_-^\dagger]}^2 = 0.
\end{equation}
and the rest of the derivation in distinguishing $W_2U_+W_1$ and $W_2U_-W_1$ simply follows. Since this holds for any $W_1, W_2$, it means that the algorithm can also distinguish \emph{any} $W_2U_+W_1$ and $W_2U_-W_1$ for any choice of $W_1, W_2$ in $\W$ when defining the target unitaries. However, the distinguishing task we consider here must have an additional layer of complexity. Indeed, when $W_1$ and $W_2$ are given to the learner \emph{before} learning $V$, they can then be taken into account during the learning procedure, which makes the task equivalent to distinguishing between $U_+$ and $U_-$ directly. We therefore consider a distinguishing task where $W_1$ and $W_2$ are revealed to the learner \emph{after} learning $V$. Then, the learner can use the learned $V$ to compute its overlap with $W_2U_+W_1$ and $W_2U_-W_1$ and distinguish between the two unitaries, but must learn $V$ without the knowledge of $W_1,W_2$.\\

\paragraph{Hardness of distinguishing $W_2U_+W_1$ and $W_2U_-W_1$}
For some choice of $(W_1, W_2)\in\W$ to be defined later, we want to bound the total variation distance between $p_+$ and $p_-$, where
\begin{equation}
    p_\pm(\psi_{1:T},\phi_{1:T})
\end{equation}
is the joint probability distribution of measuring the projections
\begin{equation}
    \kb{\phi_1}, \ldots, \kb{\phi_T} \in \{\kb{\phi_{1,j}}\}_j \times \ldots \times \{\kb{\phi_{T,j}}\}_j
\end{equation}
on the states
\begin{equation}
    W_2 U_\pm W_1\ket{\psi_1}, \ldots, W_2 U_\pm W_1\ket{\psi_T}
\end{equation}
respectively, where the states $\ket{\psi_i}$ and the POVMs $F_i = \{w_{i,j}2^n\kb{\phi_{i,j}}\}_j$ have been chosen adaptively\footnote{For conciseness, we do not index each $\psi_i/F_i/\phi_i$ by the previous choices and measurement outcomes, although these in general depend on $\psi_1, \ldots, \psi_{i-1}$, $F_1, \ldots, F_{i-1}$, and $\phi_1, \ldots, \phi_{i-1}$.}.\\

We start by considering the total variation distance of individual measurements (where $t= 1,\ldots, T$ indexes the measurement):
\begin{equation}
	\TV(p_+^t, p_-^t) = \frac{1}{2} \sum_{j} \abs{p_+(\psi_{t},\phi_{t,j}) - p_-(\psi_{t},\phi_{t,j})}
\end{equation}
and we want to show that for an overwhelming fraction of $(W_1, W_2)\in\W$, this total variation distance is exponentially small in $n$.

From a similar derivation to that of our previous Lemma, we get:
\begin{equation}\label{eq:tvd-t}
\TV(p_+^t, p_-^t) = 
\begin{cases}
\sum_{j} w_{t,j}2^n \abs{\Re[\bra{\phi_{t,j}}W_2^\dagger X_n W_1\kb{\psi_t}W_1^\dagger Z_n W_2\ket{\phi_{t,j}}]} \quad\text{for }n\text{ odd}\\
\sum_{j} w_{t,j}2^n \abs{\Im[\bra{\phi_{t,j}}W_2^\dagger X_n W_1\kb{\psi_t}W_1^\dagger Z_n W_2\ket{\phi_{t,j}}]} \quad\text{for }n\text{ even}\\
\end{cases}
\end{equation}
We now average this total variation distance over the random choice of $W_1,W_2$ (sampled uniformly from $\W$), in the case where $n$ is odd:
\begin{equation}\label{eq:avg-tvd-t}
    \E_{W_1,W_2}[\TV(p_+^t, p_-^t)] = \sum_{j} w_{t,j}2^n \E_{W_1,W_2} \left[\abs{\Re[\bra{\phi_{t,j}}W_2^\dagger X_n W_1\kb{\psi_t}W_1^\dagger Z_n W_2\ket{\phi_{t,j}}]}\right]
\end{equation}
and want to show that the expected values in this sum are all exponentially small in $n$.\\
The derivation is independent of $t,j$, so we call $X(W_1,W_2) = \Re[\bra{\phi_{t,j}}W_2^\dagger X_n W_1\kb{\psi_t}W_1^\dagger Z_n W_2\ket{\phi_{t,j}}]$ for an arbitrary $t,j$. From Jensen's inequality $g(\E[Y])\leq\E[g(Y)]$ applied to $g: x \mapsto x^2$ and $Y = \abs{X}$, we get:
\begin{align}
    \E_{W_1,W_2}[\abs{X(W_1,W_2)}] &\leq \sqrt{\E_{W_1,W_2}[X(W_1,W_2)^2]}\nonumber\\
    &= \sqrt{\E_{W_1,W_2}\left[\Re[\bra{\phi_{t,j}}W_2^\dagger X_n W_1\kb{\psi_t}W_1^\dagger Z_n W_2\ket{\phi_{t,j}}]^2\right]}.\label{eq:ineq-exp-abs}
\end{align}
From here, we make use of $\Re[z]^2 = \left(\frac{z+\overline{z}}{2}\right)^2=\frac{1}{4}(z^2+\overline{z^2}+2z\overline{z})$ for $z = \bra{\phi_{t,j}}W_2^\dagger X_n W_1\kb{\psi_t}W_1^\dagger Z_n W_2\ket{\phi_{t,j}}$ to decompose this new expected value into
\begin{equation}\label{eq:exp-sum}
    \frac{1}{4}(\E_{W_1,W_2}[z^2] + \E_{W_1,W_2}[\overline{z^2}] + 2\E_{W_1,W_2}[z\overline{z}]).
\end{equation}
The tensor product structure of $\ket{\psi_t}, \ket{\phi_{t,j}}, W_1, W_2$ and the following relation for a single-qubit unitary $W$ sampled from a 2-design and an arbitrary single-qubit state $\ket{x}$ (Eq.~(2.26) in \cite{roberts2017chaos}):
\begin{equation}
    \E_{W}\left[\left(W\kb{x}W^\dagger\right)^{\otimes 2}\right] = \frac{I + \SWAP}{6}
\end{equation}
are then useful to compute
\begin{align}
    \E_{W_1,W_2}[z^2] &= \E_{W_1,W_2}\left[\Tr[W_2\kb{\phi_{t,j}} W_2^\dagger X_n W_1\kb{\psi_t}W_1^\dagger Z_n]^2 \right]\nonumber\\
    &= \prod_{i=1}^{n} \E_{W_1^{(i)},W_2^{(i)}}\left[\Tr[W_2^{(i)}\kb{\phi_{t,j}^{(i)}} W_2^{(i)\dagger} X W_1^{(i)}\kb{\psi_t^{(i)}}W_1^{(i)\dagger} Z]^2\right]\nonumber\\
    &= \prod_{i=1}^{n} \E_{W_1^{(i)},W_2^{(i)}}\left[\Tr\left[\left(W_2^{(i)}\kb{\phi_{t,j}^{(i)}} W_2^{(i)\dagger} X W_1^{(i)}\kb{\psi_t^{(i)}}W_1^{(i)\dagger} Z\right)^{\otimes 2}\right]\right]\nonumber\\
    &= \prod_{i=1}^{n} \Tr\left[\frac{I+\SWAP}{6} X^{\otimes 2} \frac{I+\SWAP}{6} Z^{\otimes 2}\right]\nonumber\\
    &= \prod_{i=1}^{n} \frac{1}{36} (\Tr[X^{\otimes 2}Z^{\otimes 2}] + \Tr[\SWAP X^{\otimes 2}Z^{\otimes 2}] + \Tr[ X^{\otimes 2}\SWAP Z^{\otimes 2}] + \Tr[\SWAP X^{\otimes 2}\SWAP Z^{\otimes 2}])\nonumber\\
    &= \prod_{i=1}^{n} \frac{1}{36} (2\Tr[X^{\otimes 2}Z^{\otimes 2}] + 2\Tr[\SWAP X^{\otimes 2}Z^{\otimes 2}])\nonumber\\
    &= \prod_{i=1}^{n} \frac{1}{36}(0 - 4) = \left(-\frac{1}{9}\right)^n, \label{eq:exp-z2}
\end{align}
which is also the value of $\E_{W_1,W_2}[\overline{z^2}]$ since it is real. Here, we also made use of $(A\otimes B)\SWAP = \SWAP (B \otimes A)$ (since $\SWAP (A\!\otimes\!B)\SWAP = (B\!\otimes\!A)$) and $\Tr[\SWAP A\otimes B] = \Tr[AB]$.\\

Similarly, by making use of $\overline{\Tr[A]} = \Tr[\ \overline{A}\ ] = \Tr[A^\dagger]$, we have
\begin{align}
    \E_{W_1,W_2}[z\overline{z}] &= \E_{W_1,W_2}\left[\Tr[W_2\kb{\phi_{t,j}} W_2^\dagger X_n W_1\kb{\psi_t}W_1^\dagger Z_n]\Tr[W_2\kb{\phi_{t,j}} W_2^\dagger Z_n W_1\kb{\psi_t}W_1^\dagger X_n] \right]\nonumber\\
    &= \prod_{i=1}^{n} \E_{W_1^{(i)},W_2^{(i)}}\left[\Tr\left[\left(W_2^{(i)}\kb{\phi_{t,j}^{(i)}} W_2^{(i)\dagger}\right)^{\otimes 2} (X\!\otimes\!Z) \left(W_1^{(i)}\kb{\psi_t^{(i)}}W_1^{(i)\dagger}\right)^{\otimes 2} (Z\otimes X)\right]\right]\nonumber\\
    &= \prod_{i=1}^{n} \Tr\left[\frac{I+\SWAP}{6} (X\!\otimes\!Z) \frac{I+\SWAP}{6} (Z\!\otimes\!X)\right]\nonumber\\
    &= \prod_{i=1}^{n} \frac{1}{36} (\Tr[XZ\!\otimes\!ZX] + \Tr[\SWAP (XZ\!\otimes\!ZX)] + \Tr[(X\!\otimes\!Z)\SWAP (Z\!\otimes\!X)] \nonumber\\
    &\quad\quad\quad\quad\quad\quad\quad\quad\quad\quad\quad\quad\quad\quad\quad\quad\quad\quad\quad\quad\quad\quad\quad\quad + \Tr[\SWAP (X\!\otimes\!Z)\SWAP (Z\!\otimes\!X)])\nonumber\\
    &= \prod_{i=1}^{n} \frac{1}{36} (\Tr[XZ\!\otimes\!ZX] + \Tr[\SWAP (XZ\!\otimes\!ZX)] + \Tr[\SWAP] + \Tr[I\!\otimes\!I])\nonumber\\
    &= \prod_{i=1}^{n} \frac{1}{36}(0 + 2 + 2 + 4) = \left(\frac{2}{9}\right)^n \label{eq:exp-zzb}
\end{align}
Combining Eqs.~(\ref{eq:ineq-exp-abs}), (\ref{eq:exp-sum}), (\ref{eq:exp-z2}), and (\ref{eq:exp-zzb}), we get:
\begin{equation}
    \E_{W_1,W_2}[\abs{X(W_1,W_2)}] \leq \frac{1}{2}\sqrt{2\left(\frac{2}{9}\right)^n - 2\left(\frac{1}{9}\right)^n} \leq \left(\sqrt{\frac{2}{9}}\right)^n
\end{equation}
since we assumed $n$ to be odd. A similar derivation can be done for $n$ even, using $\Im[z]^2 = \left(\frac{z-\overline{z}}{2i}\right)^2=-\frac{1}{4}(z^2+\overline{z^2}-2z\overline{z})$, which leads to the same upper bound.\\
Now combining this bound with Eqs.~(\ref{eq:tvd-t}) and (\ref{eq:avg-tvd-t}), we get:
\begin{equation}\label{eq:bound-avg-tvd-t}
    \E_{W_1,W_2}[\TV(p_+^t, p_-^t)] \leq \sum_{j} w_{t,j}2^n \left(\sqrt{\frac{2}{9}}\right)^n \leq \left(\sqrt{\frac{8}{9}}\right)^n,
\end{equation}
where we made use of $\sum_{j} w_{t,j} = 1$.

Given that $\TV(p_+^t, p_-^t)$ is positive and exponentially small with respect to $n$ on average over all assignments of $(W_1, W_2)\in\W$, then, if we call:
\begin{equation}
	G(\psi_t,\phi_t) = \left\{ (W_1,W_2)\in\W : \TV(p_+^t, p_-^t) \leq \left(\frac{8}{9}\right)^{n/4} \right\},
\end{equation}
we have necessarily that
\begin{equation}\label{eq:frac-good-W}
	\frac{\abs{G(\psi_t,\phi_t)}}{\abs{\W}} \geq 1-\left(\frac{8}{9}\right)^{n/4}.
\end{equation}
We prove this by contradiction. Assuming that Eq.~(\ref{eq:frac-good-W}) does not hold, we have at least $\left(\frac{8}{9}\right)^{n/4}\abs{\W}$ assignments of $(W_1, W_2)\in\W$ for which $\TV(p_+^t, p_-^t) > \left(\frac{8}{9}\right)^{n/4}$, and therefore:
\begin{equation}
	\E_{W_1,W_2}[\TV(p_+^t, p_-^t)] = \frac{1}{\abs{\W}} \sum_{W_1,W_2 \in \W} \TV(p_+^t, p_-^t) > \left(\frac{8}{9}\right)^{n/4}\cdot \left(\frac{8}{9}\right)^{n/4} = \left(\frac{8}{9}\right)^{n/2}
\end{equation}
which contradicts with Eq.~(\ref{eq:bound-avg-tvd-t}).

From here, if we consider an entire sequence of measurements rather than a single measurement, we find that the resulting set of assignments for which $\TV(p_+^t, p_-^t)$ is exponentially small along the sequence:
\begin{equation}
	G(\psi_{1:T},\phi_{1:T}) = \left\{ (W_1,W_2)\in\W : \TV(p_+^t, p_-^t) \leq \left(\frac{8}{9}\right)^{n/4} \forall t \in 1, \ldots, T\right\}
\end{equation}
still has an overwhelming large size, for sub-exponential $T$:
\begin{equation}\label{eq:frac-good-W-seq}
	\frac{\abs{G(\psi_{1:T},\phi_{1:T})}}{\abs{\W}} \geq 1-T\left(\frac{8}{9}\right)^{n/4}.
\end{equation}
This follows immediately from the lower bound (\ref{eq:frac-good-W}) on the size of $G(\psi_t,\phi_t)$ for each $(\psi_t,\phi_t)$ along the sequence.

All we have left to do now is relate the total variation distance $\TV(p_+, p_-)$ to that of single measurements in a sequence, i.e., $\TV(p_+^t, p_-^t)$, for $t =1, \ldots, T$. For that, we show inductively that:
\begin{equation}\label{eq:tv-sequence}
	1-\TV(p_+, p_-) \geq \min_{\psi_{1:T-1}, \phi_{1:T-1}} \prod_{t=1}^T (1-\TV(p_+^t, p_-^t)).
\end{equation}
We start from the first measurement:
\begin{align}
	1-\TV(p_+, p_-) &= \sum_{\psi_{1:T}, \phi_{1:T}} \min(p_+(\psi_{1:T},\phi_{1:T}),p_-(\psi_{1:T},\phi_{1:T}))\\
	&= \sum_{\psi_1,\phi_1} \sum_{\psi_{2:T}, \phi_{2:T}} \min(p_+(\psi_1,\phi_1)p_+(\psi_{2:T},\phi_{2:T}),p_-(\psi_1,\phi_1)p_-(\psi_{2:T},\phi_{2:T}))\\
	&\geq \sum_{\psi_1,\phi_1}\min(p_+(\psi_1,\phi_1),p_-(\psi_1,\phi_1)) \sum_{\psi_{2:T}, \phi_{2:T}} \min(p_+(\psi_{2:T},\phi_{2:T}),p_-(\psi_{2:T},\phi_{2:T}))\\
		&\geq \sum_{\psi_1,\phi_1}\min(p_+(\psi_1,\phi_1),p_-(\psi_1,\phi_1)) \min_{\psi_1, \phi_1} (1-\TV(p_+^{2:T},p_-^{2:T}))\\
	&\geq (1-\TV(p_+^1,p_-^1)) \min_{\psi_1, \phi_1} (1-\TV(p_+^{2:T},p_-^{2:T}))
\end{align}
where we used $\TV(p,q) = \sum_{i, p_i \geq q_i} p_i - q_i = \sum_i p_i - \min(p_i,q_i) = 1 - \sum_i \min(p_i,q_i)$ for the first equality and the last two inequalities, and $\min(ab,cd) \geq \min(ac)\min(bd)$ for positive $a,b,c,d$ in the first inequality.

We apply the same inductive reasoning for $t=2, \ldots, T$ to get Eq.~(\ref{eq:tv-sequence}).
Since Eq.~(\ref{eq:frac-good-W-seq}) applies to any sequence, we then have from Eq.~(\ref{eq:tv-sequence}) that there exists a $(1-T\left(\frac{8}{9}\right)^{n/4})$-fraction of $\W$ for which
\begin{align}
	1-\TV(p_+, p_-) &\geq \left(1-\left(\frac{8}{9}\right)^{n/4}\right)^T\\
	&\geq 1-T\left(\frac{8}{9}\right)^{n/4}
\end{align}
which is equivalent to
\begin{equation}
\TV(p_+, p_-) \leq T\left(\frac{8}{9}\right)^{n/4} \leq T/2^{0.04n},
\end{equation}
and therefore $T$ must be scaling as $2^{\Omega(n)}$ for the algorithm to have high probability of success on all possible choices of $(W_1,W_2)\in\W$.
\end{proof}

\medskip

%% file: paper.bbl
\begin{thebibliography}{53}
\bibitem [{1()}]{1}%
  \BibitemOpen
  1,\ \href@noop {} {}\bibinfo {note} {We note that a similar argument was made
  in passing in Ref.~\cite {caro2022outofdistribution}, but the argument was
  not formalized and its implications for incoherent learning were not
  appreciated.}\BibitemShut {Stop}%
\end{thebibliography}
